\documentclass[epj,nopacs]{svjour}
\usepackage{epsfig,cite}
\usepackage{amsmath}
\usepackage{amssymb}
\usepackage{hyperref}

\def\beq{\begin{equation}}
\def\eeq{\end{equation}}
\def\ba{\begin{eqnarray}}
\def\ea{\end{eqnarray}}

\def\ss{\scriptscriptstyle}
\def\gev{{\rm \, Ge\kern-0.125em V}}
\def\tev{{\rm \, Te\kern-0.125em V}}
\def\gyr{{\rm \, G\kern-0.125em yr}}
\def\ohsq{\Omega_{\chi} h^2}

\def\nl{\hfill\nonumber\\&&}
\def\nnl{\hfill\nonumber\\}

\def\Toprel#1\over#2{\mathrel{\mathop{#2}\limits^{#1}}}

\def\sl{{\widetilde \ell}_{\scriptscriptstyle\rm R}}
\def\stau{\widetilde \tau}
\def\stop{\widetilde t}

\def\m12{m_{1\!/2}}

\def\mxi{m_{\tilde{\chi}_i^+}}

\def\mz{m_{\ss Z}}

\def\stau{\tilde{\tau}}

\def\bea{\begin{eqnarray}}
\def\eea{\end{eqnarray}}

\def\mgut{M_{GUT}}
\def\calh{\mathcal{H}}
\def\BSig{B_{\Sigma}}
\def\mSig{m_{\Sigma}}
\def\muS{\mu_{\Sigma}}
\def\hfiv{h_\mathbf{\overline{5}}}
\def\hten{h_\mathbf{10}}
\def\mfiv{m_\mathbf{\overline{5}}}

\def\mten{m_\mathbf{10}}

\def\afiv{A_\mathbf{\overline{5}}}
\def\aten{A_\mathbf{10}}
\def\alam{A_{\lambda}}
\def\alamp{A_{\lambda^\prime}}

\mathchardef\mhyphen="2D

\DeclareMathOperator{\Tr}{Tr}

\begin{document}

\title{Relating the CMSSM and SUGRA models with GUT scale and Super-GUT scale 
Supersymmetry Breaking}
\author{ Emilian Dudas\inst{1,2,3} \and Yann Mambrini\inst{3} \and 
    Azar Mustafayev\inst{4}
       \and Keith~A.~Olive\inst{4}}
\institute{Department of Physics, Theory Division, CH-1211,Geneva 23, Switzerland \and
CPhT, Ecole Polytechnique, 91128 Palaiseau, France  \and
Laboratoire de Physique Th\'eorique 
Universit\'e Paris-Sud, F-91405 Orsay, France 
       \and William I.~Fine Theoretical Physics Institute, \\
       School of Physics and Astronomy,
            University of Minnesota, Minneapolis, MN 55455, USA}
\date{Received: date / Revised version: date}

\authorrunning{Dudas et al.}
\titlerunning{Relating the CMSSM and SUGRA models}

\abstract{
While the constrained minimal supersymmetric standard model
(CMSSM) with universal gaugino masses, $m_{1/2}$, scalar masses, $m_0$, 
and A-terms, $A_0$, defined at some high
energy scale (usually taken to be the GUT scale) is motivated by general features of
supergravity models, it does not carry all of the constraints imposed by minimal supergravity
(mSUGRA).  In particular, the CMSSM does not impose a relation between the trilinear and bilinear
soft supersymmetry breaking terms, $B_0 = A_0 - m_0$, nor does it impose the relation 
between the soft scalar masses and the gravitino mass, $m_0 = m_{3/2}$. 
As a consequence, $\tan \beta$ is computed given values of the other CMSSM input parameters.
By considering a Giudice-Masiero (GM) extension to mSUGRA, one can introduce
new parameters to the K\"ahler potential which are associated 
with the Higgs sector and recover many of the standard CMSSM predictions. 
However, depending on the value of $A_0$, 
one may have a gravitino or a neutralino dark matter candidate. 
We also consider the consequences of imposing the universality conditions above the GUT scale.
This GM extension provides a natural UV completion for the CMSSM.
\vskip -2in
\begin{center}
UMN--TH--3103/12, 
FTPI--MINN--12/17, 
LPT--Orsay-12-49
\end{center}
}

\maketitle

\section{Introduction}

One of the most commonly studied variants of the minimal supersymmetric standard model
is the constrained model (CMSSM) \cite{cmssm,efgos}.  This is in part due to its simplicity 
(it is specified by four parameters), and its connection to supergravity~\cite{Fetal,acn,bfs}. 
The CMSSM also provides a natural dark matter candidate \cite{eoss}, the neutralino,
for which the relic density may be brought into the range specified by
WMAP \cite{wmap} relatively easily.  Furthermore, these models generally predict a relatively light
mass for the Higgs boson ($m_h \lesssim 130$~GeV)~\cite{ENOS}. Not only is the theory testable, 
but is currently under scrutiny from the ongoing experiments at the LHC~\cite{LHC},
resulting in strong constraints on the CMSSM parameter space, particularly when recent
constraints from Higgs searches~\cite{LHCHiggs} are applied~\cite{post-mh}.

The CMSSM is defined by choosing universal soft supersymmetry breaking parameters
input at the grand unified (GUT) scale, i.e., the scale at which gauge coupling
unification occurs. These are the universal gaugino mass, $m_{1/2}$, scalar mass, $m_0$,
and trilinear term, $A_0$. The motivation of this universality stems from minimal supergravity
(mSUGRA) and indeed the two theories are often confused. 

Minimal supergravity is defined by a K\"ahler potential with minimal kinetic terms
(in Planck units) \footnote{There are various usages of mSUGRA in the literature. 
Often mSUGRA is used as another name for the CMSSM. 
We follow the original definition of mSUGRA from Ref. \cite{acn,bfs} based on a flat K\"ahler metric
which is clearly distinct from the CMSSM. 
More general K\"ahler potentials or SUGRA models which preserve flavour symmetries are possible. Though
these are also termed mSUGRA models in the literature, they necessarily involve 
additional parameters (such as the GM model discussed below).},
\beq
G = K(\phi^i,{\phi_{i}}^*,z^{\alpha},z_{\alpha}^*) +  \ln ( |W|^2) \, ,
\label{min}
\eeq
with
\beq
K = K_0 =\phi^i {\phi_{i}}^*  + z^{\alpha} z_{\alpha}^* \, ,
\label{mink}
\eeq
where $W = f(z^{\alpha}) + g(\phi^i)$ is the superpotential,
assumed to be separable in hidden sector fields, $z^{\alpha}$, and
standard model fields, $\phi^i$. The scalar potential can be derived once
the superpotential is specified.  Assuming that the origin of supersymmetry breaking
lies in the hidden sector, the low energy potential is derived from
\begin{eqnarray}
V &=& e^{ K}\left({ K}^{I \bar J} D_I W \bar D_{\bar J}\bar{ W} -3 | W|^2\right)
\nonumber
\\
&=&e^G \left( G_I G^{I \bar J}G_{\bar J} -3 \right),
\label{eqn:SUGRApotential}
\end{eqnarray}
with $D_I W\equiv\partial_I W+{ K}_I W$
and dropping terms inversely proportional to the Planck mass, we can write~\cite{bfs}
\begin{eqnarray}
V  & =  &  \left|{\partial g \over \partial \phi^i}\right|^2 +
\left( A_0 g^{(3)} + B_0 g^{(2)} + h.c.\right)  + m_{3/2}^2 \phi^i \phi_i^*  ,
\label{pot}
\end{eqnarray}
where $g^{(3)}$ is the part of the superpotential cubic in fields,
and $g^{(2)}$ is the part of the superpotential quadratic in fields.
The trilinear term is given by
\beq
A_0  g^{(3)} = \left(\phi^i  \frac{\partial g^{(3)}}{\partial \phi^i} - 3  g^{(3)} \right) m_{3/2} + K_{\alpha} \overline{ D_{\alpha } f} (\bar{z})  g^{(3)} \ .
\label{aterm}
\eeq
Note that for trilinears, the first term in Eq. (\ref{aterm}) vanishes,
leaving
\beq
A_0 =  K_{\alpha} \overline{ D_{\alpha } f} (\bar{z})   \, ,
\eeq
while for bilinears (B-terms - defined in Eq. (\ref{aterm}) with the replacement $g^{(3)} \to g^{(2)}$), 
it is $-m_{3/2}$ yielding the familiar K\"ahler-flat
supergravity relation $B_0 = A_0 - m_0$.
In Eq. (\ref{pot}), 
the gravitino mass is given by
\beq
m_{3/2}^2 = e^G \, ,
\eeq
 and the
superpotential has been rescaled by a factor $e^{- \langle z z^* \rangle/2}$.
Finally, gaugino mass universality stems from a choice of a gauge kinetic term
which is of the form $h^A_{\alpha \beta} =h(z) \delta_{\alpha \beta}$.

Soft terms for matter fields in supergravity have a nice geometrical structure. For F-term SUSY breaking, they are given by \cite{ks}
\ba && m^2_{i \bar{j}} = m_{3/2}^2~( G_{i \bar{j}} -
R_{i\bar{j}\alpha\bar{\beta}} G^\alpha G^{\bar{\beta}} ~) \ , \nonumber \\
&& (B \ \mu)_{ij} = m_{3/2}^2~ (2  \nabla_i G_j + G^\alpha \nabla_i
\nabla_j G_\alpha ) \ , \nonumber \\
&& (A \ y)_{ijk}= m_{3/2}^2 \left( 3 \nabla_i \nabla_j G_k + G^\alpha
\nabla_i \nabla_j \nabla_k G_\alpha  \right) \ , \nonumber \\
&& \mu_{ij} = m_{3/2} \ \nabla_i G_j  \ , \nonumber \\
&& m_{1/2}^A = { 1 \over 2} ( Re \ h_A)^{-1} m_{3/2} \ \partial_{\alpha}
h_{A} \ G^\alpha \ , \label{dis1} \ea
where $y_{ijk}$ are Yukawa couplings, $h_A$ are the gauge kinetic functions and
$\nabla_i$ denotes K\"ahler covariant derivatives
\beq
\nabla_i G_j = \partial_i G_j - \Gamma_{ij}^k G_k \, ,
\eeq
where 
\beq
\Gamma_{ij}^k = G^{k \bar l} \partial_i G_{j \bar l} \, ,
 \eeq
 is the K\"ahler connection. 
$R_{i\bar{j}\alpha\bar{\beta}}$ is the Riemann tensor of the K\"ahler space spanned by chiral (super)fields. 
Taking into account the
known string compactifications, there is no reason to believe that 
they are given by very simple or even flavor universal
expressions. In order to make contact with low-energy phenomenology and in the absence of a complete viable string theory
model, one is forced, however, to resort to simplifying assumptions, for example, 
minimal supergravity as defined in Eq.~(\ref{mink}).

In the CMSSM, however, it is customary to drop the mSUGRA relation
between $B_0$ and $A_0$.
Instead, $B_0$ and the Higgs mass mixing term, $\mu$, are solved
using the low energy electroweak symmetry breaking conditions,
i.e., from the minimization of the Higgs potential at $M_{weak}$.
Furthermore, in the CMSSM, the relation between $m_0$ and the gravitino mass is
lost, though scalar mass universality is maintained. 
As a results, phenomenological constraints in the CMSSM can be displayed
on a $(m_{1/2}, m_0)$ plane, for fixed $A_0$ and $\tan \beta$. Note the sign of the $\mu$
parameter must also be specified. In contrast, in mSUGRA models,
because of the relation between $B_0$ and $A_0$, $\tan \beta$ is no longer a free parameter
\cite{vcmssm}, and we are left with three free parameters (rather than four).

An interesting extension of minimal supergravity
is one where terms proportional $g^{(2)}$ are added to the K\"ahler
potential as in the Giudice-Masiero mechanism~\cite{gm}. 
For example, consider an additional contribution to $K$,
\beq
\Delta K = c_H H_1 H_2  + h.c. \, ,
\label{gmk}
\eeq
where $c_H$ is a constant, and $H_{1,2}$ are the usual MSSM
Higgs doublets.  
Notice that in string theory $c_H < 1$ is needed for the viability of the effective field theory limit~\cite{smallc}. 
The effect of $\Delta K$, is manifest on the boundary conditions for 
both $\mu$ and the $B$ term at the supersymmetry breaking input scale, $M_{in}$.
The $\mu$ term is shifted to
\beq
\mu + c_H m_0 \, .
\eeq
Note that while in principle we can define an input value for $\mu$ ($\mu_0$), 
it is not determined by supersymmetry breaking and furthermore, 
since we solve for $\mu$ at the weak scale, its UV value is fixed by the low energy
boundary condition.
The boundary condition on $\mu B$ shifts from $\mu B_0$
to 
\beq
\mu B_0 + 2 c_H m_0^2 \, .
\eeq
It is clear therefore, that using the GM mechanism,
one can avoid altogether a dimensionful quantity in the superpotential
(i.e., one can set $g^{(2)} = 0$) and obtain a weak scale $\mu$ proportional 
to $c_H m_0$. While the extension in Eq. (\ref{gmk}) is perhaps the simplest
extension which affects the $B$-term, it is by no means unique.
However, the GM extension is the simplest
mechanism to solve the $\mu$-problem, 
that plagues SUGRA realizations of the MSSM.

In principle, we can also use $\Delta K$ to better connect the CMSSM
to supergravity. Indeed, by allowing $c_H \ne 0$, we can once again
fix $\tan \beta$ and derive $\mu$ and $B \mu$ at the weak scale. 
The presence of the extra term in the K\"ahler potential allows one
to match the supergravity boundary conditions at $\mgut$.
In particular, by running our derived value of $B(M_W)$ up to the GUT scale,
we can write
\beq
B(\mgut) = (A_0 - m_0) + 2 c_H m_0^2/\mu(\mgut) \, .
\label{gmb1}
\eeq
Indeed, we can use Eq.~(\ref{gmb1}) to derive the necessary value of $c_H$.
So long as $c_H \lesssim O(1)$, we can associate the CMSSM with
this non-minimal version of supergravity which we will refer to as GM supergravity.

For numerical computations we employed the program {\tt SSARD}~\cite{ssard}, which uses 2-loop RGE evolution for the MSSM and
1-loop evolution for minimal SU(5) to compute the sparticle spectrum. 
These are passed to {\tt FeynHiggs}~\cite{FeynHiggs} for computation of the light Higgs boson mass, $m_h$.
Throughout this paper we take the top quark mass $m_t = 173.1$~GeV~\cite{mt} and 
the running bottom quark mass $m_b^{\overline{MS}}(m_b) = 4.2$~GeV~\cite{rpp}.

In section \ref{sec:GUT}, we consider this connection between the CMSSM and 
GM supergravity.  In particular, we will show that for essentially all
CMSSM models of interest, the values of $c_H$ are small enough to remain in the perturbative
regime.  We next consider a super-GUT version of the CMSSM based on minimal SU(5)
for which the supersymmetry breaking input scale is increased above $\mgut$~\cite{emo,Calibbi}.
We first demonstrate that in the context of mSUGRA, the standard boundary conditions
for the $B$-term are very difficult to satisfy and generally require that the coupling, 
$\lambda$ between the Higgs five-plets and the Higgs adjoint is small (close to 0). 
This is similar to what was
found for a no-scale supergravity GUT~\cite{nosc1}. Generally the no-scale sparticle
spectrum is problematic unless one moves the input supersymmetry breaking scale above
$\mgut$~\cite{ENO}. As a consequence, strong constraints can be derived on the coupling 
$\lambda$~\cite{emo2}. In section \ref{sec:zerocSig}, we will show the effect of turning on the coupling
$c_H$ (now defined as a coefficient of the five-plets, $\calh_{1}\calh_2$).
In this case, CMSSM-like planes can be defined, albeit with strong constraints on the 
coupling $\lambda$. That is, while the boundary conditions can be matched, the resulting
solution for $c_H$ becomes wildly non-perturbative.
 In section~\ref{sec:nonzcSig}, we show that these constraints can relaxed
 if we turn on an additional contribution to the K\"alher potential, namely
 $c_\Sigma Tr \Sigma^2 + h.c.$, which can be associated with the $\mu$ and $B$ terms
 of the Higgs adjoint.  This will in principle, lead to a family of solutions relating
 $c_H$ and $c_\Sigma$. Our conclusions will be given in section~\ref{sec:concl}.

\section{GUT Scale Universality}
\label{sec:GUT}

We will begin by exploring the bridge between the CMSSM and mSUGRA
via an addition to the K\"ahler potential when the input supersymmetry breaking
scale is $M_{in} = \mgut$.
The addition to the K\"ahler potential can be chosen as
given in Eq.~(\ref{gmk}).
In the CMSSM, $\mu$ and $B$ are normally solved for 
in terms of $m_Z$ and $\tan \beta$:
\ba
\mu^2 & = & \frac{m_1^2 - m_2^2 \tan^2 \beta + \frac{1}{2} \mz^2 (1 -
\tan^2 \beta) + \Delta_\mu^{(1)}}{\tan^2 \beta - 1 + \Delta_\mu^{(2)}} \, ,
\nonumber \\
B \mu  & = & -{1 \over 2} (m_1^2  + m_2^2 + 2 \mu^2) \sin 2 \beta + \Delta_B \, ,
\label{onelooprel}
\ea
where $\Delta_B$ and $\Delta_\mu^{(1,2)}$ are loop corrections~\cite{Barger:1993gh,deBoer:1994he,Carena:2001fw}, 
and  $m_{1,2}$ are the Higgs soft masses (here evaluated at the weak scale). 
As a result, there is usually a one-to-one correspondence between $B$ and $\tan \beta$,
so that there is perhaps a single value for $\tan \beta$ for which 
the GUT-scale\footnote{The GUT scale, $\mgut$, is defined as the scale where SU(2) and U(1) gauge couplings
unify and is approfimately $1.5\times 10^{16}$~GeV.}
 boundary condition, $B_0 = A_0 - m_0$ is satisfied. 

\begin{figure*}
\begin{center}
\epsfig{file=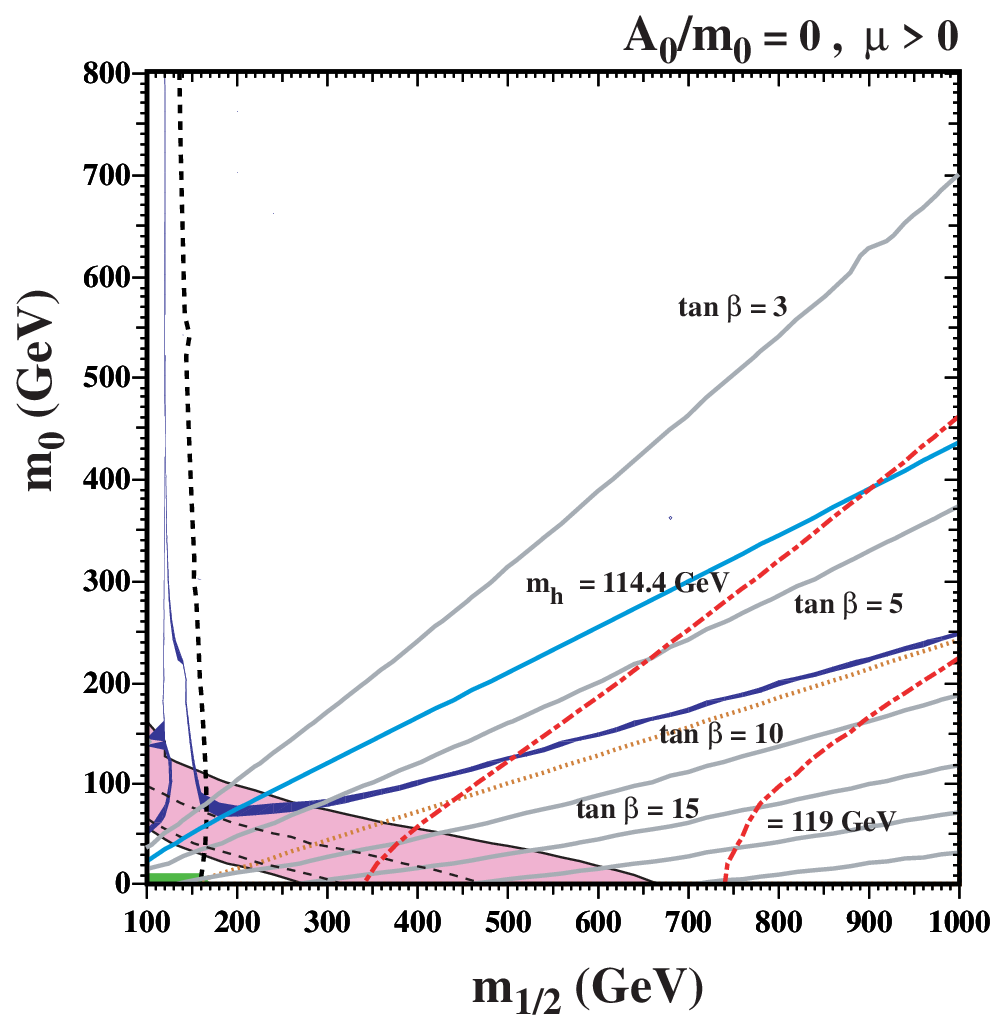,height=8.5cm}
\epsfig{file=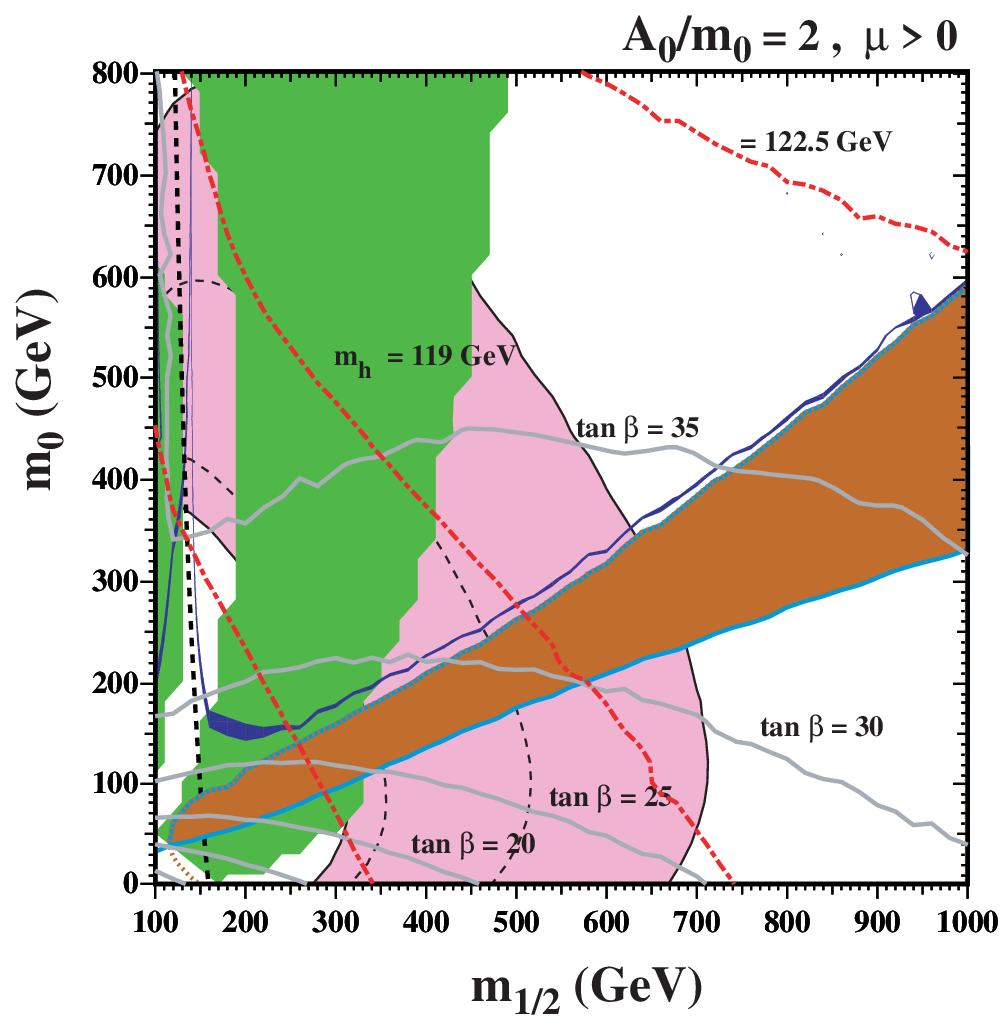,height=8.5cm}\\
\end{center}
\caption{\it
The $(m_{1/2}, m_0)$ planes for minimal supergravity model with 
$A_0/m_0 = 0$ (left) and  $A_0/m_0 = 2$(right). 
The relic density is within the WMAP range in the blue strip. The
pink region between the black dashed (solid) lines is allowed by $g_\mu-2$ at 
the 1-$\sigma$ (2-$\sigma$) level. The gravitino is the LSP below the diagonal light blue line and
$m_{\stau_1} < m_\chi$ below the brown dotted curve.
The brown and green colored regions are excluded by the requirements of a neutral LSP, and
by $b \to s \gamma$, respectively. The contours for $m_h$ are labeled in the figure
and are shown as red dot-dashed curves and the contours for $\tan \beta$ are shown as
solid gray curves. The black dashed curve is the $\mxi = 104$ GeV contour.
More details can be found in the text.}
\label{fig:msugra}
\end{figure*}

We show in Fig.~\ref{fig:msugra} the allowed parameter space in a $(m_{1/2},m_0)$ plane
for mSUGRA
with $A_0=0$ (left) and $A_0=2m_0$ (right) (updated from Ref.~\cite{vcmssm}). 
Here, and in subsequent
figures, the regions forbidden because the lightest supersymmetric particle (LSP) 
is charged (either ${\tilde \tau_1}$ or ${\tilde t_1}$) are
shaded brown, the regions excluded by $b \to s \gamma$~\cite{bsg} are shaded
green, the regions favored by $g_\mu - 2$~\cite{newBNL} at the $\pm 2 - \sigma$ level are shaded
pale pink, with the $\pm 1-\sigma$ region bordered by dashed curves. 
The near vertical black dashed line is the chargino mass $m_{\chi^\pm_1} = 104$~GeV contour and the 
red dot-dashed lines show contours of the Higgs mass, $m_h$ as labelled. 
Unlike the CMSSM, each point on the plane corresponds to a value of $\tan \beta$
and these are shown by the gray-colored curves for $\tan \beta = 3$ and 
in increments of 5 (most are labeled on the figure).
For $A_0/m_0 = 0$, much of the plane at large $m_0$ has small $\tan \beta \lesssim 5$ and 
a correspondingly small value of $m_h$. For $A_0/m_0 = 2$, higher values of $\tan \beta$ are found
and they extend up to $\sim 39$ in the region plotted.  

The dark blue shading in Fig.~\ref{fig:msugra} 
indicates the region where the relic density falls within the WMAP range, 
$0.097 \leq \Omega_{CDM} h^2 \leq 0.122$. 
We also plotted the limit $M_{LSP}=m_{3/2}$ shown as the light blue diagonal line
under which the gravitino is the LSP. It corresponds roughly to the line $m_0=0.4 m_{1/2}$.
Another diagonal line (brown dotted) shows the contour for which the lightest neutralino mass 
$m_\chi$ is equal to the mass of the lighter stau,  $m_{\tilde \tau_1}$.
For $A_0/m_0 = 0$, the latter appears below the gravitino LSP line, and as such, ${\tilde \tau_1}$
is never the LSP. As a consequence, only the dark blue shaded region
at low $m_{1/2}$ above the light blue line corresponds to neutralino dark matter at the WMAP
density. The dark blue shaded region below the light blue line corresponds to the 
gravitino LSP at the WMAP density assuming that there is no nonthermal 
contribution to the gravitino density (valid for example in models where the inflationary
reheat temperature is rather low). Here, the gravitino density is determined
from the relic annihilations of either the neutralino or stau (if below the dotted line)
and $\Omega_{3/2} h^2 = (m_{3/2}/m_{\chi,{\tilde \tau_1}}) \Omega_{\chi,{\tilde \tau_1}} h^2$.
However, in regions with a gravitino LSP, there are additional constraints 
from big bang nucleosynthesis (not considered here) which may impact its viability~\cite{bbn}.
This is in fact a conservative constraint as the gravitino relic density maybe higher if
a large abundance of gravitinos are produced during reheating after inflation.

As shown previously \cite{vcmssm}, one observes that an extended region respecting the WMAP
relic density with a neutralino LSP appears for larger values of $A_0$ as a result of stau
 coannihilation~\cite{stauco} as seen in the right panel of Fig.~\ref{fig:msugra}.
 Indeed, for large values of the trilinear coupling, the mass of the lighter stau, $\tilde \tau_1$,
is lower which pushes the coannihilation channel to regions of the parameter space
where $m_{\chi_0} \simeq m_{\tilde \tau_1} > m_0 = m_{3/2}$ . In this case, it is even possible
to satisfy WMAP with a relatively heavy Higgs ($m_h \gtrsim 122$~GeV for $\tan\beta \gtrsim 37$).
Notice in this case, below the co-annihilation strip, there is a region (as in the CMSSM)
where  $\tilde \tau_1$ is the LSP and hence shaded brown.  At still lower $m_0$,
the gravitino is once again the LSP with a $\tilde \tau_1$ being the next to lightest supersymmetric
particle (NLSP). Note also, that the region excluded by $b \to s \gamma$ (shaded green) is
significantly more important than the case with small $A_0$. In fact, for $A_0/m_0 = 2$,
we see that the excluded region is split.
This occurs because BR($b \to s \gamma$) is too large at small $m_{1/2}$, falls through
the acceptable range as $m_{1/2}$ increases, becoming unacceptably small because of
cancellations over a range of $m_{1/2}$, before rising towards the Standard Model value
at large $m_{1/2}$.

\begin{figure*}
\begin{center}
\epsfig{file=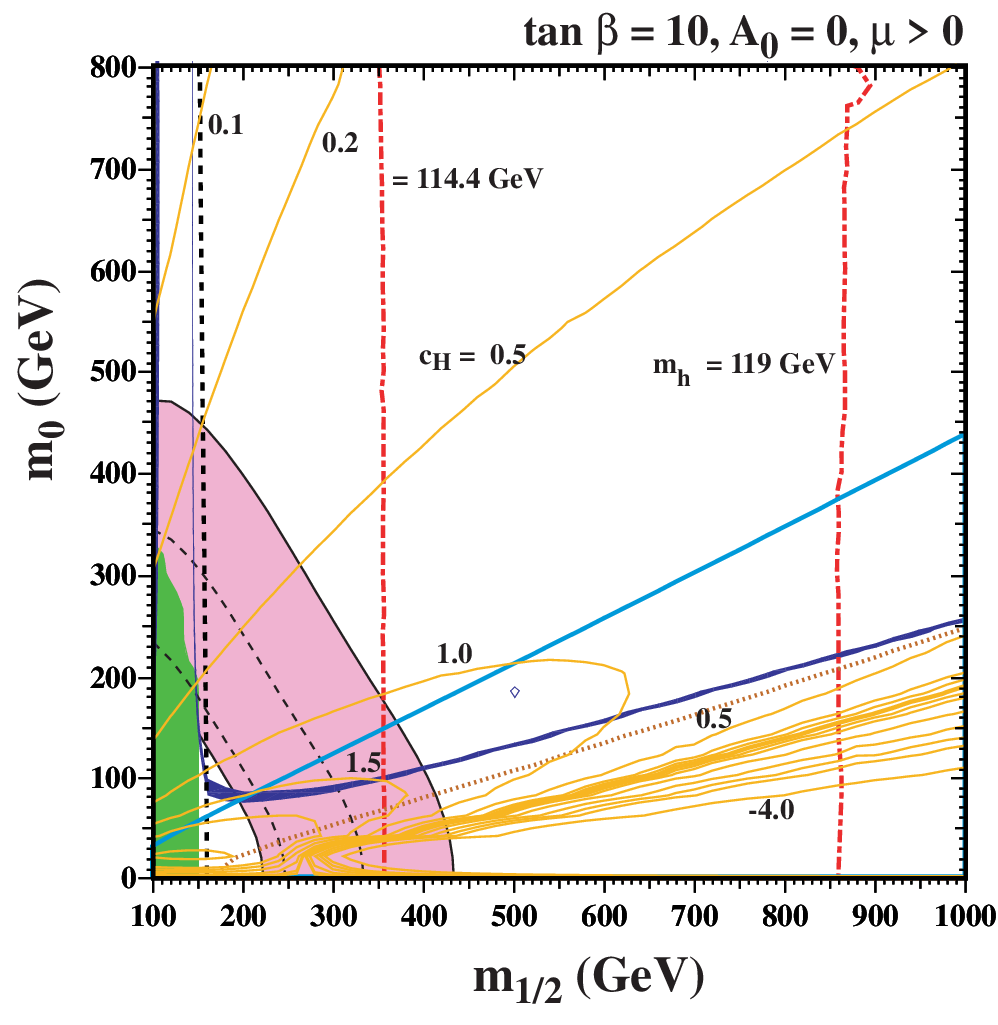,height=8.5cm}
\epsfig{file=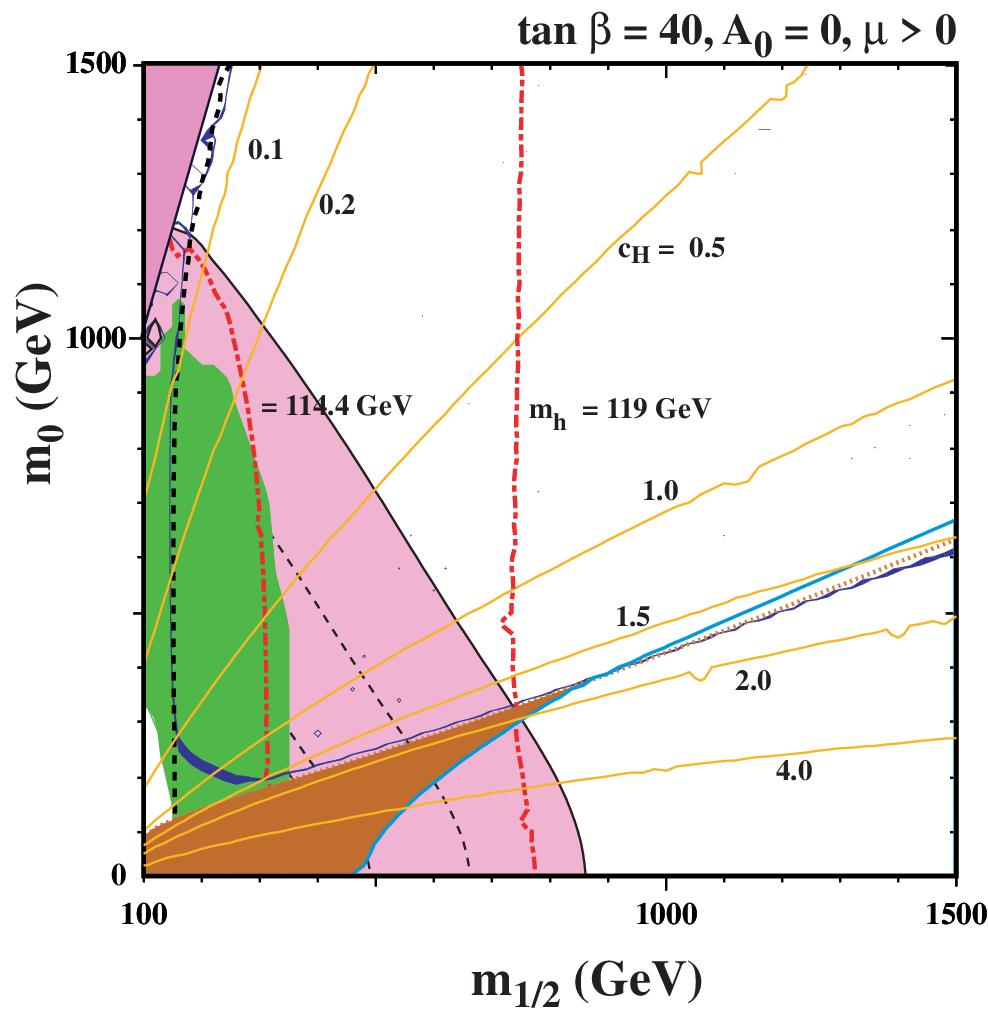,height=8.5cm}\\
\end{center}
\caption{\it
The $(m_{1/2}, m_0)$ planes for CMSSM based on a  GM supergravity model 
with $A_0 = 0$, $c_H \ne 0$ and with 
$\tan \beta = 10$ (left) and  $\tan \beta = 40$(right). 
The meaning of the curves and shaded regions are the same as in Fig.~\ref{fig:msugra}.
However, here we show the contours of the required value of $c_H$ in order to maintain
the fixed value of $\tan \beta$ across the plane.
For $\tan \beta = 40$,  it is not possible to satisfy
the electroweak symmetry breaking conditions in the 
the dark pink shaded region at low $m_{1/2}$ and high $m_0$.}
\label{fig:gmsugra0}
\end{figure*}

When the Giudice-Masiero term (\ref{gmk}) is included \cite{gm}, 
 one can deduce the (GUT) boundary conditions for $\mu$ and $B$
\bea
&&
\mu = \mu_0 +  c_H m_0 \, ,
\\
&&
B_0 = A_0-m_0+2 c_H m_0^2/\mu_0 \, .
\label{gmb}
\eea
Of course the first of these is irrelevant as we still solve for $\mu$ at the weak scale
using (\ref{onelooprel}) and $\mu_0$ is arbitrary. However, Eq.~(\ref{gmb}) now allows one
to solve for $B$ at the weak scale for an arbitrary $\tan \beta$, and still satisfy the
GUT scale supergravity boundary condition, thus solving for $c_H$. 
Therefore, relaxing the condition between $A_0$ and $B_0$ and considering
 $\tan \beta$ as an input, as is done in the CMSSM,
  is equivalent to ``switching on" the coefficient $c_H$ in Eq.~(\ref{gmb}).
In other words, for a given value of $\tan \beta$ and $A_0$, at  
each point ($m_{1/2}$,$m_{0}$) there may exist a single value of $c_H$ respecting Eq.~(\ref{gmb}).
We display the iso-$c_H$ contours in Fig.~\ref{fig:gmsugra0} for $A_0=0$ and $ \tan \beta=10$ and 40.

The $(m_{1/2},m_0)$ planes shown in Fig.~\ref{fig:gmsugra0} resemble
standard CMSSM planes~\cite{eoss} as recently updated in~\cite{eo6}.
The first remarkable result seen in these figures, is the ``natural" values of $c_H$ that 
one obtains in the region
of parameter space of interest: $0.1 \lesssim c_H \lesssim 1$.
As might be expected, values of $c_H$ become very large at small $m_0$, i.e.,
in the gravitino LSP region.  While we can forgo the relation between $B_0$ and
$\tan \beta$ in GM supergravity, we can not escape the relation $m_{3/2} = m_0$.
Thus for $\tan \beta = 10$, as seen in the left panel of Fig.~\ref{fig:gmsugra0},
the WMAP co-annihilation strip largely falls in the gravitino LSP region.
The unmarked contours of $c_H$ between 0.5 and -4 correspond to (0.2, 0.1, 0, -0.1,
-0.2, -0.5, -1, -1.5, and -2). The contour for $c_H = 0$ is slightly thicker and notice
that this corresponds exactly to the contour for $\tan \beta = 10$ in Fig.~\ref{fig:msugra}.
For $\tan \beta = 40$ as seen in the right panel of Fig.~\ref{fig:gmsugra0},
there is a co-annihilation strip between $m_{1/2} \simeq 300 - 700$~GeV which extends
to Higgs masses up to $\sim 119$~GeV. However, here, the values of $c_H \sim 1.5 -2$.
The familiar stau co-annihilation region is limited to relatively low $m_{1/2}$.
Towards the upper left of this panel, there is a region where there is no consistent electroweak vacuum 
and it is shaded (darker) pink. The thin dark blue strip following that border corresponds to the focus point
region~\cite{focus}.

\begin{figure*}
\begin{center}
\epsfig{file=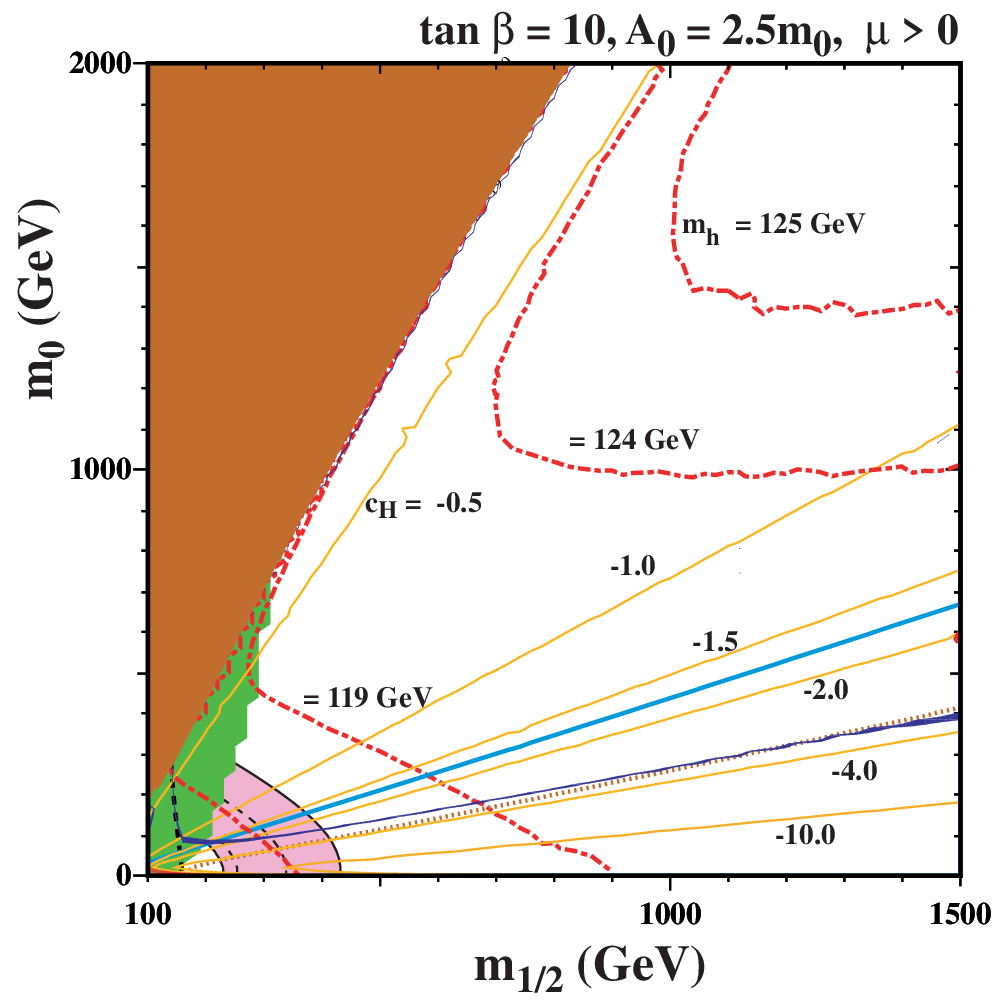,height=8.5cm}
\epsfig{file=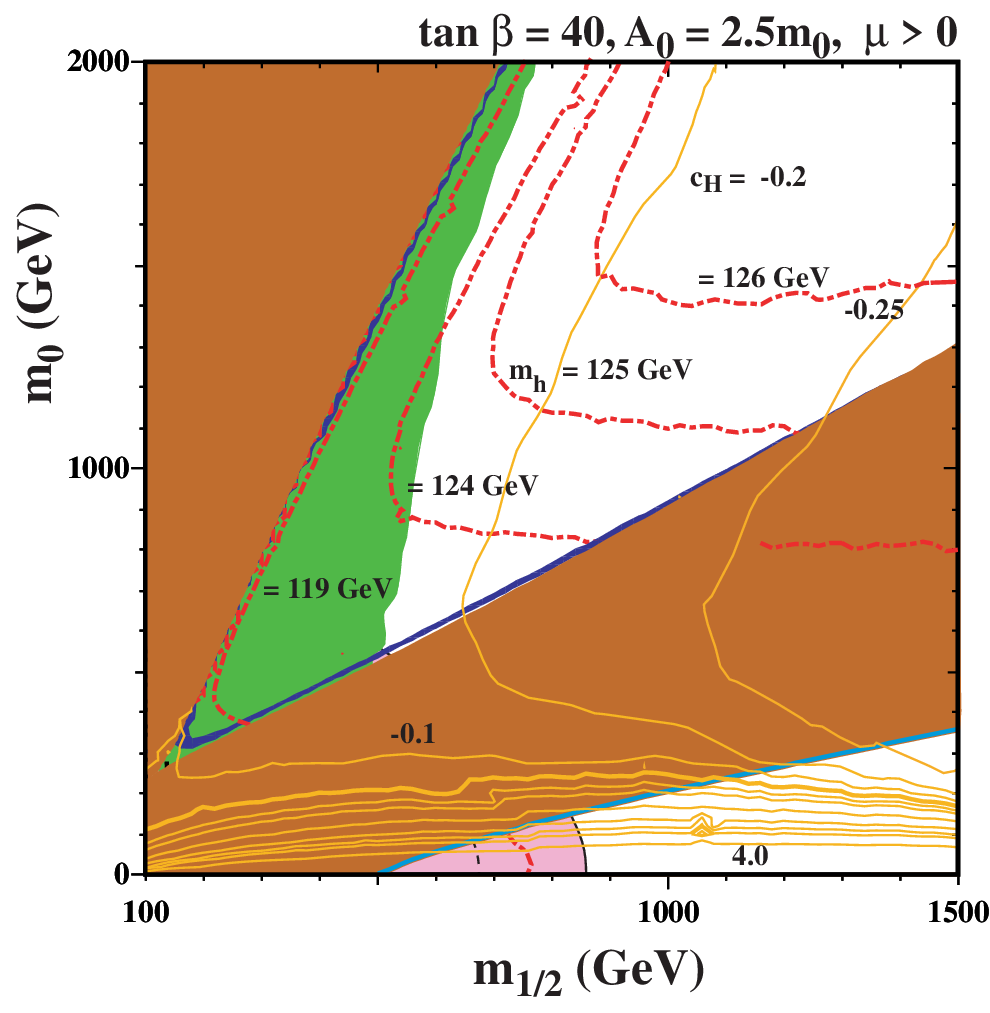,height=8.5cm}\\
\end{center}
\caption{\it
As in Fig. \ref{fig:gmsugra0} with $A_0/m_0 = 2.5$,  and
$\tan \beta = 10$ (left) and  $\tan \beta = 40$(right). The brown shaded
triangular region in the upper left of the figures has a stop LSP (or tachyonic stop). 
For $\tan \beta = 40$, there is a also a brown shaded region with a stau LSP
(above the gravitino region in the lower right).}
\label{fig:gmsugra2.5}
\end{figure*}

The parameter plane becomes even more interesting if $A_0 \neq 0$ as shown in Fig.~\ref{fig:gmsugra2.5}
for $A_0/m_0 = 2.5$ for the same two values of $\tan \beta$. In each of these panels,
we show regions shaded brown in the upper left corner corresponding to the parameter space
with a stop LSP. Though it is difficult to see, there is a stop co-annihilation~\cite{dm:stop} strip running 
along side of it, however, in the case of $\tan \beta = 40$, this strip is excluded by $b \to s \gamma$
\cite{eo6}. For $\tan \beta = 10$ the stop co-annihilation strip (with $m_h \simeq 119$~GeV), remains
viable, however, the stau co-annihilation strip, lies predominantly in the gravitino LSP region. 

For $\tan \beta = 40$, 
there exists a region of the parameter space where the model can fulfill the WMAP constraint and reach a Higgs mass of 125 GeV 
for $c_H \simeq -0.25$. 
We can easily understand why higher
values of the trilinear coupling $A_0$ leads to smaller values for the parameter $c_H$.
From Eq.~(\ref{gmb}), for a given value of $m_0$, increasing $A_0$ requires
a decrease in $c_H$ if one is to conserve the same value of $B$ at GUT scale (and thus
the same value of tan$\beta$). This is clearly illustrated by comparing 
Figs.~\ref{fig:gmsugra0} and \ref{fig:gmsugra2.5}
where, for example, the point $m_{1/2}=m_{0}=1000$~GeV needs $c_H\simeq0.6$ if $A_0=0$ and $c_H\simeq -0.25$
when $A_0=2.5 m_0$. This property of the dependence of the $c_H$ coefficients will play 
an important role when we will analyze the case $M_{in}> \mgut$.

\begin{figure*}
\begin{center}
\epsfig{file=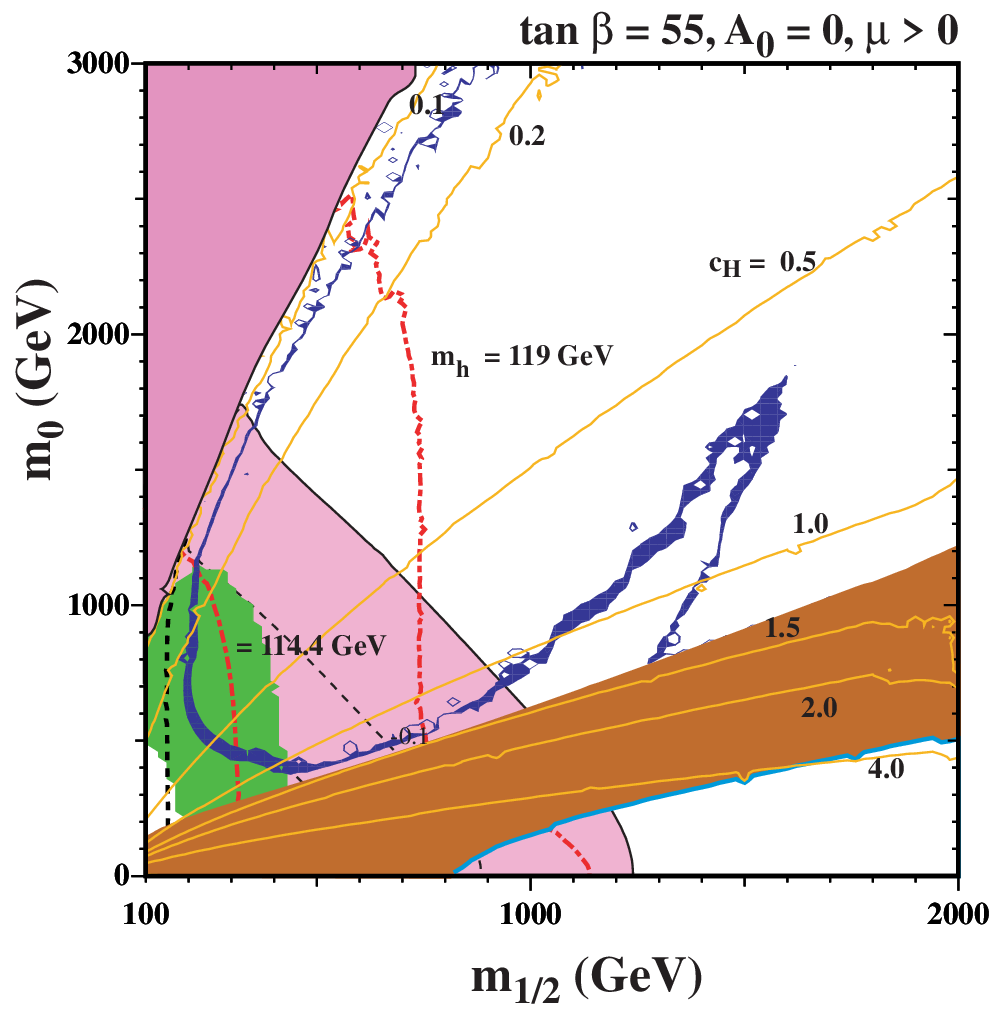,height=8.5cm}
\epsfig{file=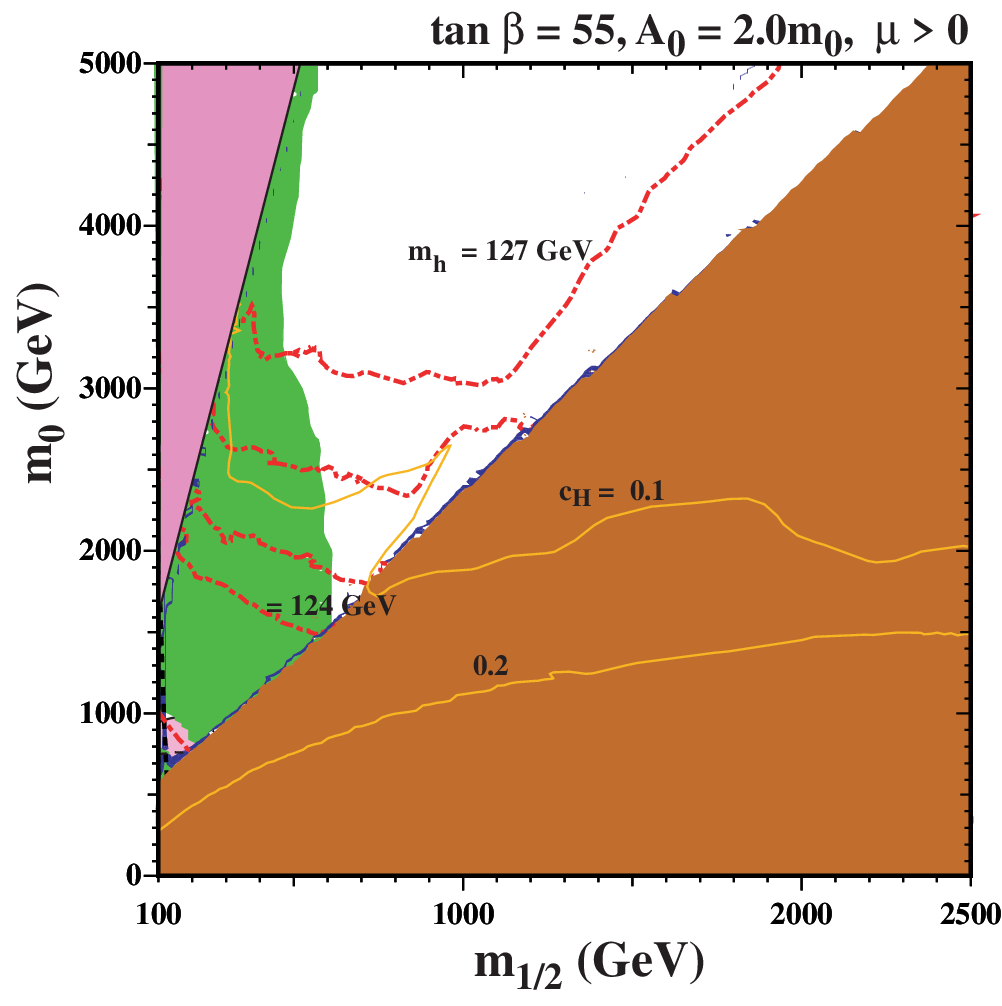,height=8.5cm}\\
\end{center}
\caption{\it
As in Fig. \ref{fig:gmsugra0} with $\tan \beta = 55$ ,  and
$A_0/m_0 = 0$ (left) and  $A_0/m_0 = 2.0$ (right). }
\label{fig:gmsugra55}
\end{figure*}

In Fig.~\ref{fig:gmsugra55}, we show analogous planes for $\tan \beta = 55$ and $A_0 = 0$
(left) and $A_0 = 2.0 m_0$ (right). For $A_0  = 0$, all of the regions with acceptable
relic density correspond to a neutralino LSP. In this case, we see the appearance
of the rapid Higgs annihilation funnel~\cite{funnel,efgos} where neutralinos annihilate primarily through
s-channel heavy Higgs exchange. As one can see, the funnel lies in an area where
$c_H < 1.5$ and the Higgs mass reaches $\sim 122$~GeV. We again see a region (in the upper left)
with no electroweak symmetry breaking and a focus point strip which tracks it near the $c_H = 0.1$
contour. 
For $A_0  = 2.0 m_0$, the $\tilde \tau_1$ being even lighter (even tachyonic for low $m_0$),
 one finds the correct relic abundance and $m_h=125$~GeV for $c_H=0.1$ (the unmarked
 Higgs mass contours in this panel correspond to 125 and 126~GeV). Notice that there is no
 gravitino LSP region for the parameters displayed.

\section{Super-GUT Scale Universality}
\label{sec:supergut}

While it is common to assume that the input supersymmetry breaking scale is equal to the GUT scale, 
it is quite plausible that $M_{in}$ may be either below~\cite{eosk} (as in models with mirage 
mediation~\cite{mixed})
or above~\cite{emo,Calibbi,pp,emo3,Baer:2000gf} the GUT scale. Here, we will consider the latter.
Increasing $M_{in}$ increases the renormalization of the soft masses which tends in turn to increase the splittings between the physical sparticle masses~\cite{pp}.  
As a consequence, the coannihilation strip is moved to lower values of $m_{1/2}$
In addition,  the focus-point strip often moves out
to very large values of $m_0$. This feature of super-GUT models is essential for 
models such as those described in Ref.~\cite{lmo} in which gaugino masses (and $A$-terms)
are produced via anomalies while scalar masses remain equal to $m_{3/2}$ at $M_{in}$,
thus requiring very large $m_0$. 

To realize $M_{in} > \mgut$, we need to work in the context of a specific GUT.
Here, we use the particle content and the 
renormalization-group equations (RGEs)
of minimal SU(5)~\cite{pp,others}, primarily for simplicity: 
for a recent review of this sample model and its
compatibility with experiment, see~\cite{Senjanovic:2009kr}.
As this specific super-GUT extension of the CMSSM was studied extensively in Refs.~\cite{emo,emo2},
we refer the reader there for details of the model.  

We note here that in our super-GUT framework, we integrate out all extra multiplets at the scale $\mgut$, 
so the theory below $\mgut$ has the same field content as in the MSSM. 
However, this differs from the CMSSM, as the RGE running above the GUT scale generates a particular non-universal pattern 
for MSSM soft terms at $\mgut$. This model then also differs from commonly studied NUHM models \cite{nuhm}, where the
non-universality is present only in the Higgs soft masses. Here, gaugino masses as well as sfermion masses
are non-universal at $\mgut$. The degree of non-universality will depend on $M_{in}$ as well as 
GUT-specific couplings. Furthermore because of the matching at $\mgut$ of the $B$-terms
(there are two in minimal SU(5)), the mSUGRA relation between the MSSM $A$ and $B$-terms will not hold
at $\mgut$ (though an analogous relation at $M_{in}$ will be valid) 
and hence the superGUT theory we describe is (in principle) distinguishable
from mSUGRA and thus have the appearance of a more general SUGRA model with non-universal soft masses. 
Thus while the mSUGRA model we describe is a subset (a one-parameter reduction) of the 
CMSSM, mSUGRA and the CMSSM are only part of the superGUT family in the limit that
$M_{in} \to \mgut$.

The model is defined by the superpotential
\ba
W_5 &=& \muS \Tr\hat{\Sigma}^2 + \frac{1}{6}\lambda'\Tr\hat{\Sigma}^3
 + \mu_H \hat{\calh}_{1} \hat{\calh}_2 
 + \lambda \hat{\calh}_{1}\hat{\Sigma} \hat{\calh}_2 \nl
 +({\bf h_{10}})_{ij} 
   \hat{\psi}_i \hat{\psi}_j \hat{\calh}_2 
 +({\bf h_{\overline{5}}})_{ij} \hat{\psi}_i \hat{\phi}_{j} \hat{\calh}_{1} \, ,
\label{W5}
\ea
where $\hat{\phi}_i$ ($\hat{\psi}_i$) correspond to the $\bf{\overline{5}}$ ($\bf{10}$) representations
of superfields, $\hat{\Sigma}(\bf{24})$, $\hat{\calh}_1(\bf{\overline{5}})$ and 
$\hat{\calh}_2(\bf{5})$ represent the Higgs adjoint and five-plets.  
Here $i,j=1..3$ are generation indices and we suppress the SU(5) index structure for brevity. 
There are now
two $\mu$-parameters, $\mu_H$ and $\muS$,  as well as two new couplings, $\lambda$
and $\lambda^\prime$.  Results are mainly sensitive to $\lambda$ and the ratio of the
two couplings. In what follows, we will fix $\lambda^\prime = 1$. 

In the context of GM supergravity, 
the K\"ahler potential can be written as
\beq
K=K_0 + c_H {\calh_1} {\calh_2} + \frac{1}{2} c_{\Sigma} \Tr \Sigma^2 + h.c. \, ,
\label{gmk2}
\eeq
where ${\calh_{1,2}}$  are scalar components of the Higgs five-plets and $\Sigma$ is the scalar component of the adjoint Higgs.
Thus in principle, we have two extra parameters which can be adjusted to relate
the CMSSM and supergravity boundary conditions for $M_{in}>\mgut$.

The breaking $SU(5)\rightarrow SU(3)_c \times SU(2)_L \times U(1)_Y$ arises 
from the Standard-Model singlet component $\hat{\Sigma}_{24}$,
that develops a vev of $\mathcal{O}(\mgut)$, \newline
$\langle \hat{\Sigma}\rangle = \langle \hat{\sigma}\rangle \,  diag(2,2,2,-3,-3)$. 
The latter can be decomposed as
\beq
\langle \hat{\sigma} \rangle = 
  \langle \sigma \rangle + \theta^2 \langle \mathcal{F}_{24} \rangle ,
\eeq
where $\sigma$ and $\mathcal{F}_{24}$ are, respectively, the scalar and auxiliary field components of 
the superfield $\hat{\sigma}$. Note that since SU(5) is broken at the scale 
$\langle \hat{\sigma} \rangle \sim \mgut$ 
and the supersymmetry breaking scale is $\sim M_{weak}$, 
the dominant contribution to the scalar component vev is 
$v_{24}=2\sqrt{30}\mu_\Sigma/\lambda'$ and is  $\mathcal{O}(\mgut)$, 
while the corresponding contribution to the auxiliary field is of the order of the weak scale.

Ignoring the couplings to matter fields, the corresponding scalar potential including soft SUSY-breaking lagrangian terms is
\bea
V({\calh_1},{\calh_2},\sigma) & = & \left|\frac{\partial W_5}{\partial {\calh_1}}\right|^2 + 
\left|\frac{\partial W_5}{\partial {\calh_2}}\right|^2 + \left|\frac{\partial W_5}{\partial \sigma}\right|^2 \nnl
&  +  & ({\Delta \mu_H}^2 + 2 \mu_H {\Delta \mu_H}+  m^2_{\calh_1}) |\calh_1|^2  \nnl
& + & ({\Delta \mu_H}^2 + 2 \mu_H {\Delta \mu_H} + m^2_{\calh_2})|\calh_2|^2   \nnl
    & +& ({\Delta \mu_\Sigma}^2 + 2 \mu_\Sigma{\Delta \mu_\Sigma} + m^2_{\Sigma}) |\sigma|^2 \nnl
	& + & \left[\frac{1}{2} b_\Sigma \sigma^2 + b_H  \calh_{1}\calh_2 
	-  \frac{1}{6\sqrt{30}}\alamp\lambda' \sigma^3
		       \right. \nl
 \left.  -\frac{1}{2} \sqrt{\frac{6}{5}} \alam\lambda\calh_{1}\calh_2 \sigma
 -\frac{\lambda^\prime}{2\sqrt{30}} \Delta \mu_H \sigma |\sigma|^2  \right. \nnl
&&  \left. -\frac{1}{2} \sqrt{\frac{6}{5}}  \mu_\Sigma \lambda\calh_{1}\sigma^* \calh_2  +h.c. \right] ,
\label{softH}
\eea
The additional terms in the K\"ahler potential (\ref{gmk2}) introduce new terms in the scalar potential that are of similar structure to those coming from
the superpotential $\mu$-terms. Therefore it is convenient to define effective $\mu$ parameters as
\beq
{\tilde \mu_\Sigma} = \mu_\Sigma + 
\Delta \mu_\Sigma \, ,
\eeq
 such that at the scale $M_{in}$, 
 \beq
 {\tilde \mu_\Sigma}(M_{in}) = \mu_\Sigma(M_{in}) + 
c_\Sigma m_0 \, ,
\label{muSigBC}
\eeq 
and similarly for $\tilde \mu_H$.  We also define an effective $b = B\mu$ term as
\beq
{\tilde b}_\Sigma = b_\Sigma +\Delta b_\Sigma \, ,
\eeq
which at $M_{in}$ is given by
\beq
{\tilde b}_\Sigma(M_{in}) = b_\Sigma(M_{in}) + 2 c_\Sigma m_0^2 \, ,
\label{bSigBC}
\eeq 
and similarly for $\tilde b_H$.

Then, at the scale $M_{in}$, we impose universal SUGRA boundary conditions
\ba
m_0 & = & \mfiv=\mten=m_{\calh_1}=m_{\calh_2}=\mSig, \nnl
A_0 & = & \afiv=\aten=\alam =\alamp , \nnl
m_{1/2} & = & M_5 \, ,
\label{sugBC}
\ea
where $M_5$ is the SU(5) gaugino mass, 
and evolve all parameters to $\mgut$ using the SU(5) RGEs.
In addition, we must impose the SUGRA relation on $B$-terms, 
\beq
 B_H = \BSig = B_0\equiv A_0 - m_0\, .
\label{superB}
\eeq
At the GUT scale, the SU(5) parameters must be matched to their MSSM counterparts.
This matching has been
studied carefully in Ref.~\cite{Borzumati}, and we make use of their results here. 
Of interest to us here, are the matching conditions for the $\mu$- and $B$-terms.

The MSSM Higgs bilinears $\mu$ and $B$ can be expressed in terms of SU(5) parameters as
\bea
\mu  &=& \mu_H -\frac{3}{\sqrt{30}}\lambda \langle\sigma\rangle = {\tilde \mu_2} + \delta \mu_2 , \nnl
b \equiv B\mu &=& b_H -\frac{3}{\sqrt{30}}\lambda 
        \left( \alam \langle\sigma\rangle + \langle\mathcal{F}_{24}\rangle  \right) \nnl
        & = &  {\tilde b_2} + \delta b_2 \, .
\label{match}
\eea
$B_2$ and $\mu_2$ (and therefore $b_2$) are the corresponding bilinears of the electroweak doublets inside the five-plets 
and are given by 
\bea 
{\tilde \mu_2} & = & {\tilde \mu_H} - 6\frac{\lambda}{\lambda'}\muS , \nonumber\\
\delta \mu_2 & = & 6 \frac{\lambda}{\lambda'}(\BSig - \alamp - \Delta \muS), \nonumber\\
{\tilde b_2} & = & {\tilde b_H} - 6\frac{\lambda}{\lambda'} \muS (\alam - \alamp + \BSig) \nnl
& = & B_H \mu_2 + \Delta b_H + 6 \frac{\lambda}{\lambda'}\mu_\Sigma \Delta , \label{b2} \nonumber\\
\delta b_2 & = & - 6 \frac{\lambda}{\lambda'} \left[(\BSig - \alamp)(\BSig - \alam)  + \mSig^2 \right. \nnl
& &  \left. \qquad + (\alam - \alamp)\Delta \muS + \Delta b_\Sigma  \right] \, ,
\label{b2mu2}
\eea
where $\mu_2  = {\tilde \mu_2} - \Delta \mu_H$. 
The quantity $\Delta\equiv B_H-\alam-\BSig-\alamp$ that appears in the third expression of (\ref{b2mu2}) 
is RGE invariant (at one loop) and it is equal to zero by universal boundary conditions (\ref{sugBC}) and (\ref{superB}).
The first of the expressions in (\ref{b2mu2}) represents the well-known
doublet-triplet fine-tuning which balances the two GUT-scale quantities, $\mu_H$ and $\muS$
to obtain the weak-scale $\mu_2$. 

Note that the MSSM parameters $\mu$ and $b$ are fixed at the weak scale
by the minimization of the Higgs potential as in the CMSSM. These quantities 
can be run up to the GUT scale using common MSSM RGEs. 
While the couplings $\lambda$ and $\lambda^\prime$ are fixed at the GUT scale,
they can be run up to $M_{in}$ so that the quantities $\delta \mu_2$ and $\delta b_2$
can be unambiguously fixed at $M_{in}$. Both of these depend on the GM parameter $c_\Sigma$. With the help of the RGE's given below,
 $\delta \mu_2$ and $\delta b_2$ can be run down to $\mgut$. At $\mgut$, the
quantities ${\tilde \mu_2}$ and ${\tilde b_2}$ are computed using expressions (\ref{match}) 
and need to be evolved back to $M_{in}$. 
At $M_{in}$, the SUGRA boundary conditions for ${\tilde b_2}$ allow us to solve for 
$c_H$ (for a given $c_\Sigma$) leading to the expression
\beq
c_H=\frac{{\tilde b_2}+(m_0-A_0){\tilde \mu_2}}{m_0(3m_0-A_0)} \, .
\label{ch}
\eeq

From their expressions (\ref{b2mu2}) 
we see\footnote{The quantity $\lambda \muS = \lambda' v_{24}$ evolves as $\mu_H$, 
hence $\mu_2$ and $b_2$ evolve as $\mu_H$ and $b_H$, respectively~\cite{Borzumati}. 
Quantities ${\tilde \mu_H}$ and ${\tilde b_H}$ evolve also as $\mu_H$ and $b_H$,  
since they are represent the same terms in the Lagrangian.}
that ${\tilde \mu_2}$ and ${\tilde b_2}$ 
evolve as $\mu_H$ and $b_H$, respectively, i.e. their RGEs are
\bea
\frac{d\tilde \mu_2}{dt} &=& \frac{1}{16\pi^2}{\tilde \mu_2}
  \left[ 48\hten^2+2\hfiv^2+\frac{48}{5}\lambda^2 -\frac{48}{5}g_5^2 \right] \, , \nnl
\frac{d \tilde b_2}{dt} &=& \frac{\tilde b_\Sigma}{16\pi^2}
  \left[ 48\hten^2+2\hfiv^2+\frac{48}{5}\lambda^2 -\frac{48}{5}g_5^2 \right] \nnl
 & + &  \frac{\tilde \mu_2}{8\pi^2}
  \left[ 48\aten \hten^2+2\afiv \hfiv^2+\frac{48}{5}\alam \lambda^2 -\frac{48}{5}g_5^2\right]   \, .
\eea
On the other hand, ${\Delta \mu_\Sigma}$ and ${\Delta \mu_\Sigma}$ are set at $M_{in}$ by 
(\ref{muSigBC}) and (\ref{bSigBC}) and need to be evolved down to $\mgut$. 
Their RGEs are the same as the ones for $\muS$ and $b_\Sigma$, respectively:
\bea
\frac{d\Delta \mu_\Sigma}{dt} &=& \frac{1}{8\pi^2}{\Delta \mu_\Sigma}
  \left[ \lambda^2+\frac{21}{20}{\lambda'}^2 -10g_5^2\right] \, , \nnl
\frac{d\Delta b_\Sigma}{dt} &=& \frac{\Delta b_\Sigma}{8\pi^2}
  \left[ \lambda^2+\frac{21}{20}{\lambda'}^2 -10g_5^2\right] \nnl
& + & \frac{\Delta \mu_\Sigma}{8\pi^2}
  \left[ 2\alam \lambda^2+\frac{42}{20}\alamp {\lambda'}^2 +20M_5 g_5^2\right] \, .
\eea
Other relevant RGE's can be found in Refs.~\cite{pp,emo,emo2}.

We now successively consider the impact of $M_{in} > \mgut$ in the context of supergravity.
We first turn off the GM terms, which leave us with an mSUGRA model with $M_{in} > \mgut$.
Next, as in the previous section, we consider the effect of $c_H \ne 0$, which will already allow
us to break the mSUGRA relation for $b_2$ as seen in Eq.~(\ref{b2}) by the additional term
$\Delta b_H$. As we will see, in some portions of the parameter space, $c_H$ is rather
large and we explore the possibility that $c_H$ can be adjusted by taking $c_\Sigma \ne 0$.

\subsection{$c_H = 0, c_\Sigma = 0$}
\label{sec:zerocH}

\begin{figure*}
\begin{center}
\epsfig{file=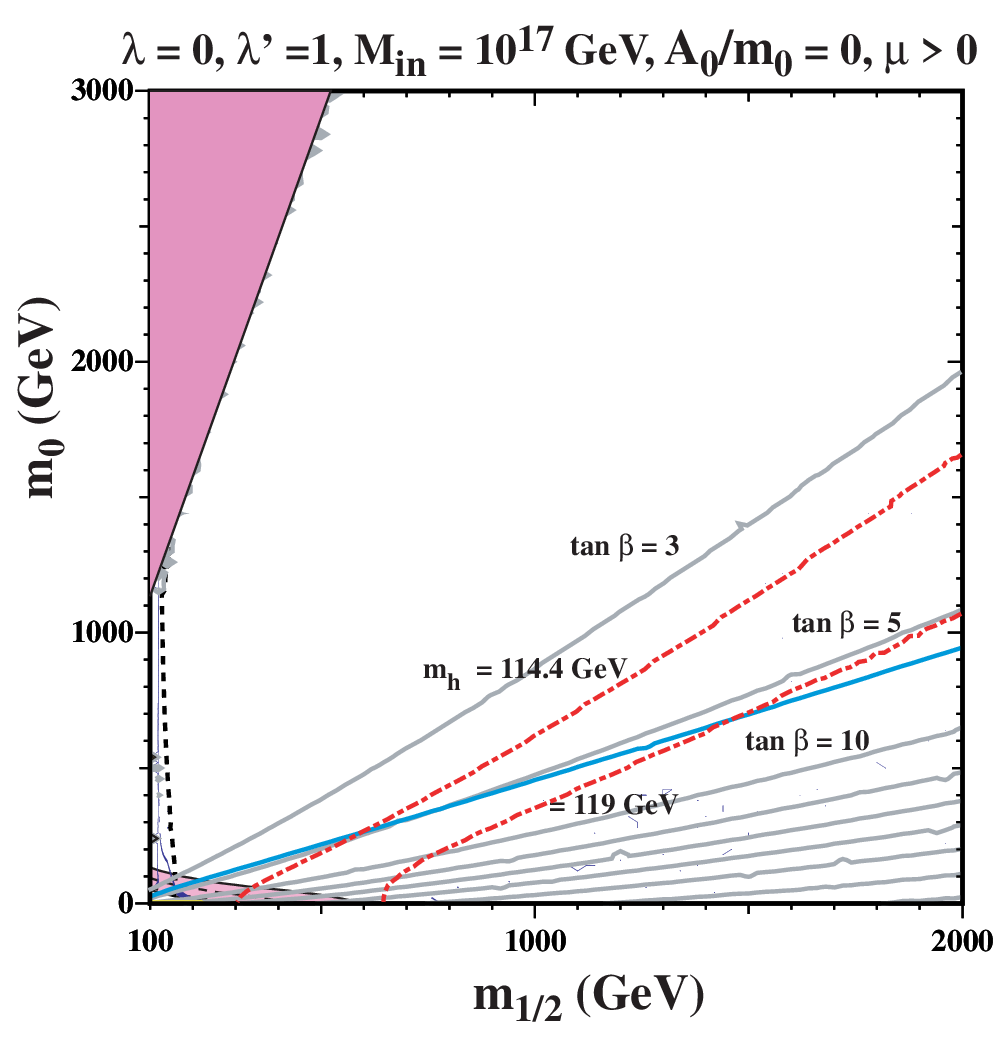,height=8.5cm}
\epsfig{file=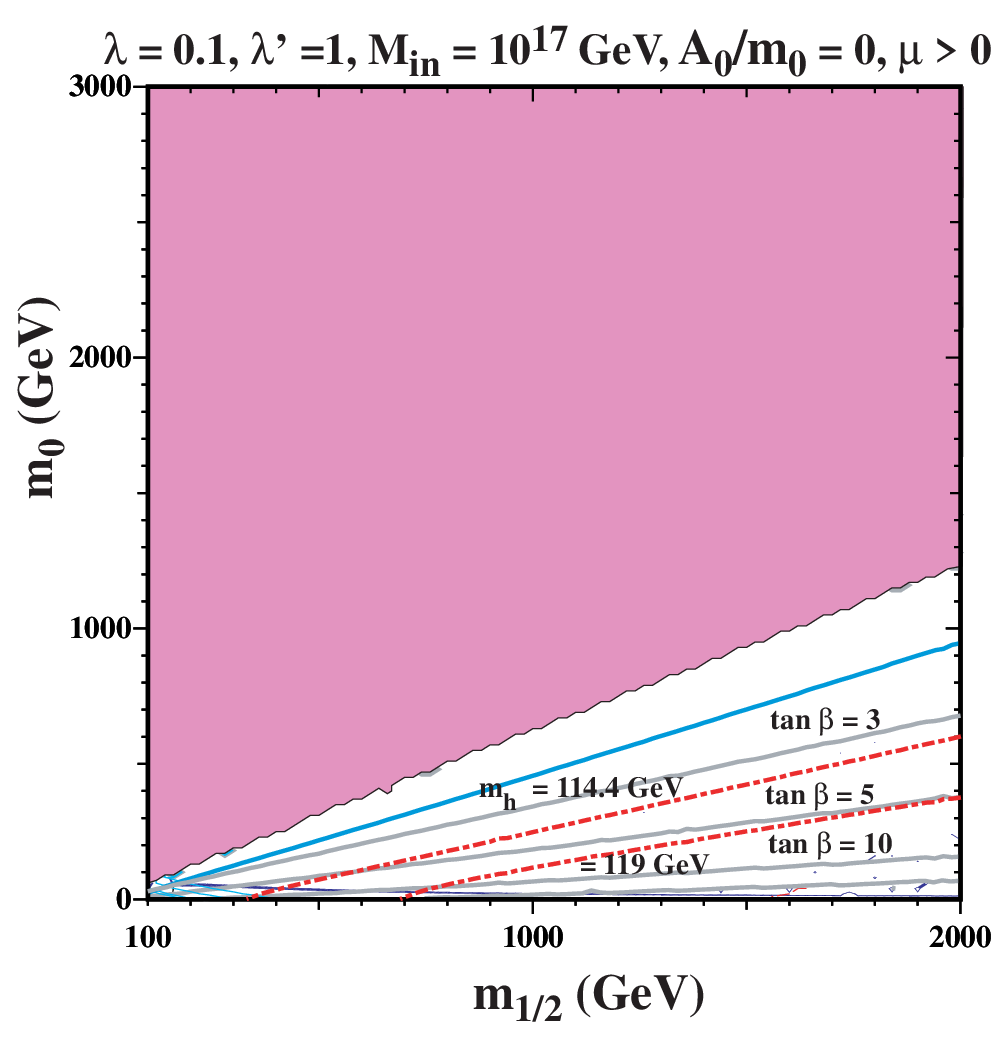,height=8.5cm}\\
\end{center}
\caption{\it
As in Fig.~\ref{fig:msugra}, but for the minimal SU(5) with  $A_0/m_0 = 0$, $M_{in} = 10^{17}$~GeV,
$\lambda^\prime = 1$, and $\lambda = 0$ (left) and  $\lambda = 0.1$ (right). }
\label{fig:msugra0-17}
\end{figure*}

As noted earlier, 
imposing the boundary conditions at $M_{in}> \mgut$ dramatically changes the picture 
of the mSUGRA model. In Fig.~\ref{fig:msugra0-17}, we show the  ($m_{1/2},m_0$) plane
for  mSUGRA with
$M_{in} = 10^{17}$~GeV, $A_0 = 0$, and $\lambda =0$ (left) and $\lambda =0.1$ (right).
This should be compared with Fig.~\ref{fig:msugra} with GUT scale universality.
The region where the $m_{\tilde \tau_1} < m_{\chi}$ has effectively disappeared.
The region where the relic density matches the WMAP determination is present
only in the lower left corner of the figure.
Once again, $\tan \beta$ is solved at each point, and we show contours of fixed
$\tan \beta$. 

However, as one can see in the right panel of 
Fig.~\ref{fig:msugra0-17}, a non-zero value for $\lambda$, even as
small as $0.1$, can almost entirely close the mSUGRA parameter space due
to the lack of solutions to the electroweak symmetry breaking conditions at 
the weak scale. 
This behavior can be understood from the b-term. For vanishing $A_0$, $\tilde b_2$ starts 
out negative at $M_{in}$.
The running of the b-term is small because the gauge and yukawa contributions are opposite and 
almost cancel each other. 
As a result, $B<0$ and small values of $\tan\beta$ are required to satisfy the 
EWSB condition (\ref{onelooprel}). 
As $\lambda$ increases, $B$ is driven more negative 
because of the increasingly negative contribution from $\delta b_2$.
This leads to quickly diminishing values of $\tan\beta$ until the EWSB condition 
can no longer be satisfied~\cite{emo2}.  
This conclusion is amplified if $M_{in}$ is increased further. Since the results are qualitatively
similar, we keep $M_{in}$ fixed at $10^{17}$~GeV.
For higher value of $\lambda$,
there are no solutions to the supergravity boundary conditions which yield
a solution for $\tan \beta$. For $\lambda \gtrsim 0.5$, the entire space (for the parameter
range shown) is closed.

\begin{figure*}
\begin{center}
\epsfig{file=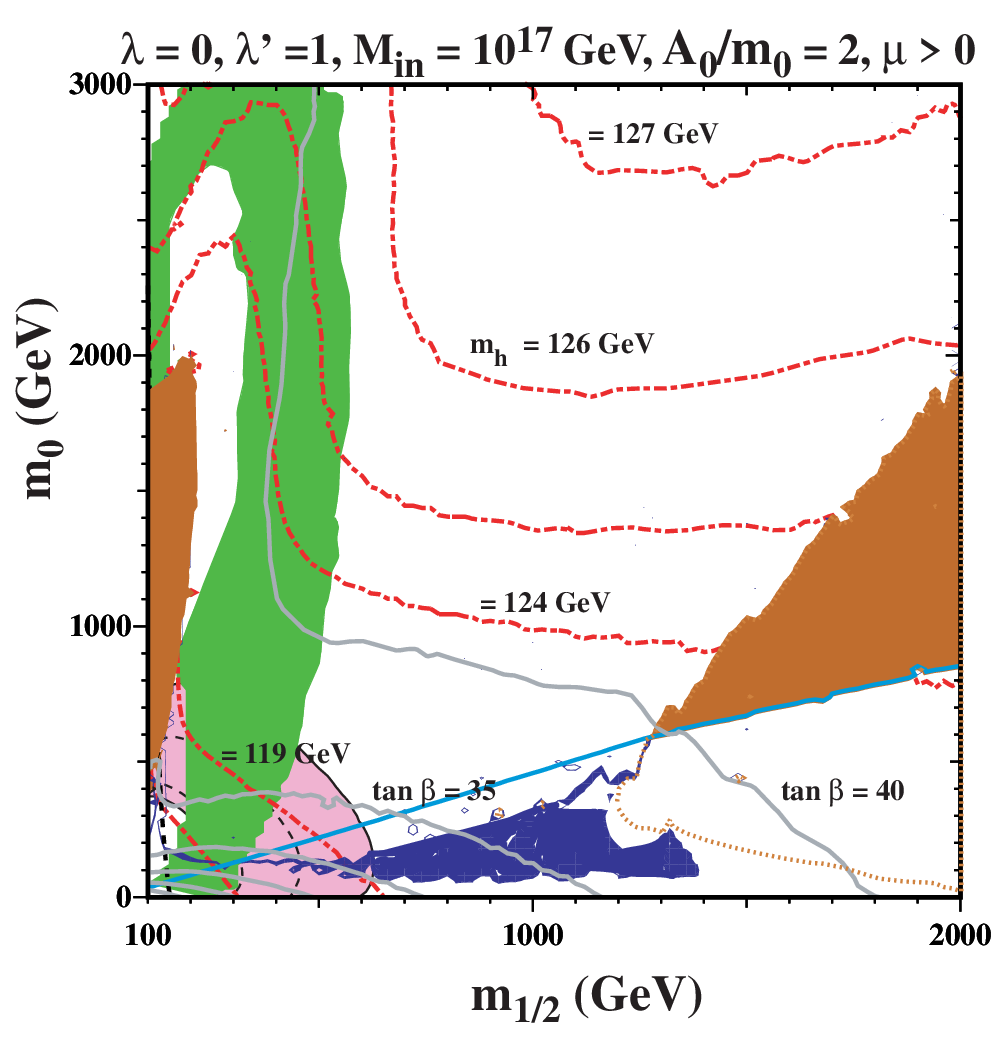,height=8.5cm}
\epsfig{file=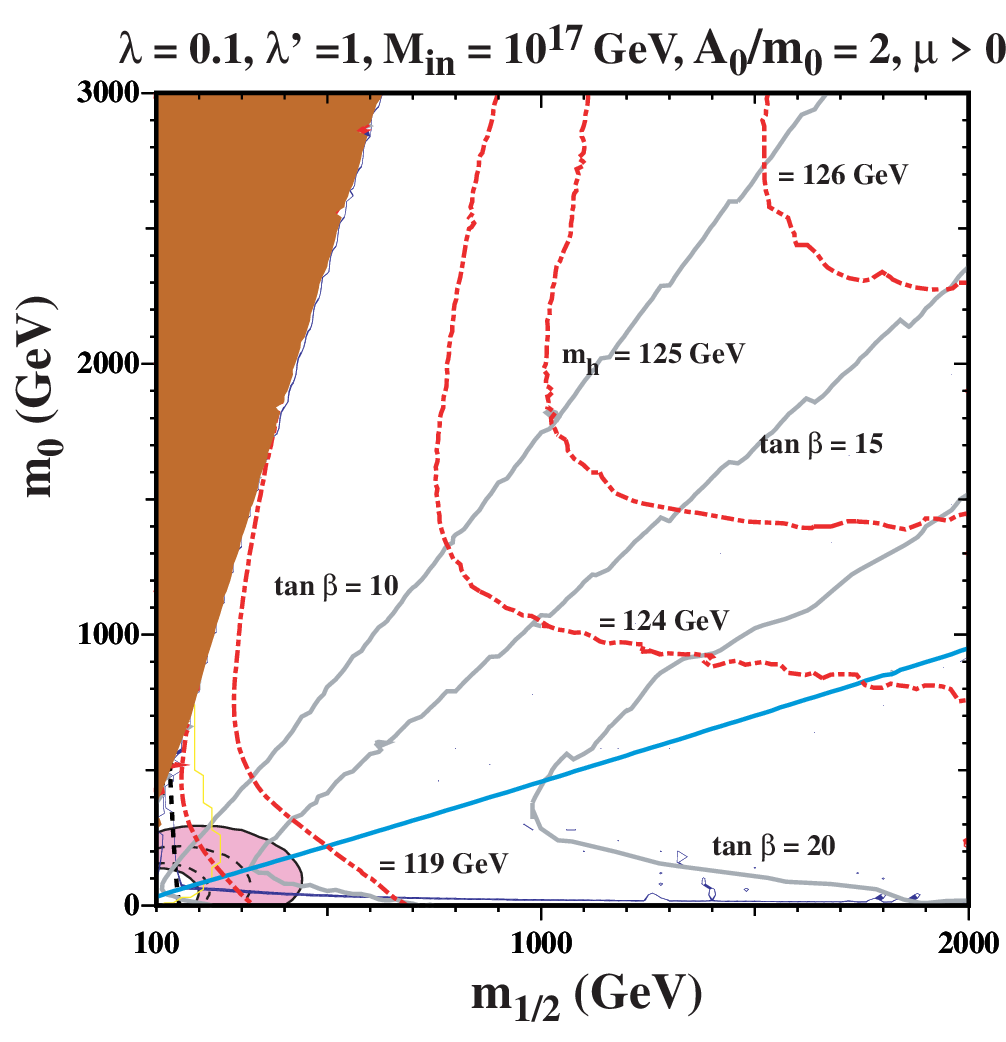,height=8.5cm}\\
\end{center}
\caption{\it
As in Fig. \ref{fig:msugra0-17} with $A_0/m_0 = 2$, again with
$\lambda = 0$ (left) and  $\lambda = 0.1$ (right). }
\label{fig:msugra2-17}
\end{figure*}

Increasing the value of $A_0$ reopens the parameter space because it increases the 
value of ${\tilde b_2}(M_{in})$, 
and, since its contribution to the RGE between $\mgut$ and $M_{weak}$ is negative, it can counterbalance 
the influence of $\lambda$. 
As seen in Fig.~\ref{fig:msugra2-17}, for $A_0=2m_0$ the EWSB condition can easily be satisfied over the entire
parameter plane. 
However, larger $A_0$ also lowers sfermion masses due to RGE effects and increased left-right mixing. 
For $\lambda = 0$,  as seen in the left panel of Fig.~\ref{fig:msugra2-17}, a region with 
 $m_{\tilde \tau_1} < m_{\chi}$ has reappeared, now to the right of plane, a portion
 of which is above the gravitino LSP line, and thus shaded brown.
 There is another brown shaded region to the left of the plane, where
 $m_{\tilde t_1} < m_{\chi}$. As in the right panel of Fig.~\ref{fig:msugra}, 
 there is also a large region where the constraint from $b \to s \gamma$ is relevant.
 As expected, values of $\tan \beta$ and $m_h$ are higher relative to the case with $A_0 = 0$.
 
When $\lambda  = 0.1$ as in the right panel of Fig.~\ref{fig:msugra2-17},
values of $\tan \beta$ are very different demonstrating the dependence on
the ratio $\lambda/\lambda^\prime$ in Eq.~(\ref{b2mu2}). In this case, the stop
co-annihilation region is pronounced and there is a region of good relic density 
which tracks along side of it with $m_h \approx 119$~GeV.
As mentioned earlier, larger $\lambda$ leads to smaller values of $\tan\beta$. 
This, in turn, lowers the $\stop_1$ mass so that the excluded stop-LSP 
region grows larger in the right panel. 
Smaller $\tan\beta$ has the opposite effect on the $\stau_1$ mass: a smaller tau Yukawa coupling produces a smaller
downward push in the $m^2_{\stau_R}$ running, and since $\stau_1 \simeq \stau_R$, $m_{\stau_1}$ becomes larger.
Hence, the stau-LSP excluded region disappears in the right panel. 
For couplings $\lambda \gtrsim 0.8$, we again lose our ability to solve for $\tan \beta$. 

As in the case of GUT-scale universality, we can in principle, restore some
of the conclusions found for the CMSSM with $M_{in} > \mgut$, 
by considering GM supergravity. In the next section, we will
analyze how these different contributions
affect the CMSSM parameter space and determine the required values of $c_H$.

\subsection{$c_H \ne 0, c_\Sigma = 0$}
\label{sec:zerocSig}

Next we show results for $M_{in} = 10^{17}$~GeV when we allow
$c_H \ne 0$.  As in the case of GUT input scale supersymmetry breaking, 
with $c_H \ne 0$, we can in principle fix $\tan \beta$ and solve for $c_H$
for any given $m_{1/2}, m_0$ and $A_0$.  
When $c_\Sigma = 0$, the expression for $\delta b_2$ takes its mSUGRA form~\cite{Borzumati}, 
and the boundary condition for $b_2$ is once again,
$b_2 (M_{in}) = (A_0 - m_0) \mu_2(M_{in}) + 2c_H m_0^2$, where
$\mu_2(M_{in})$ is determined by running $\mu_2(\mgut) = \mu(\mgut) - \delta \mu_2(\mgut)$
up to $M_{in}$.

\begin{figure*}
\begin{center}
\epsfig{file=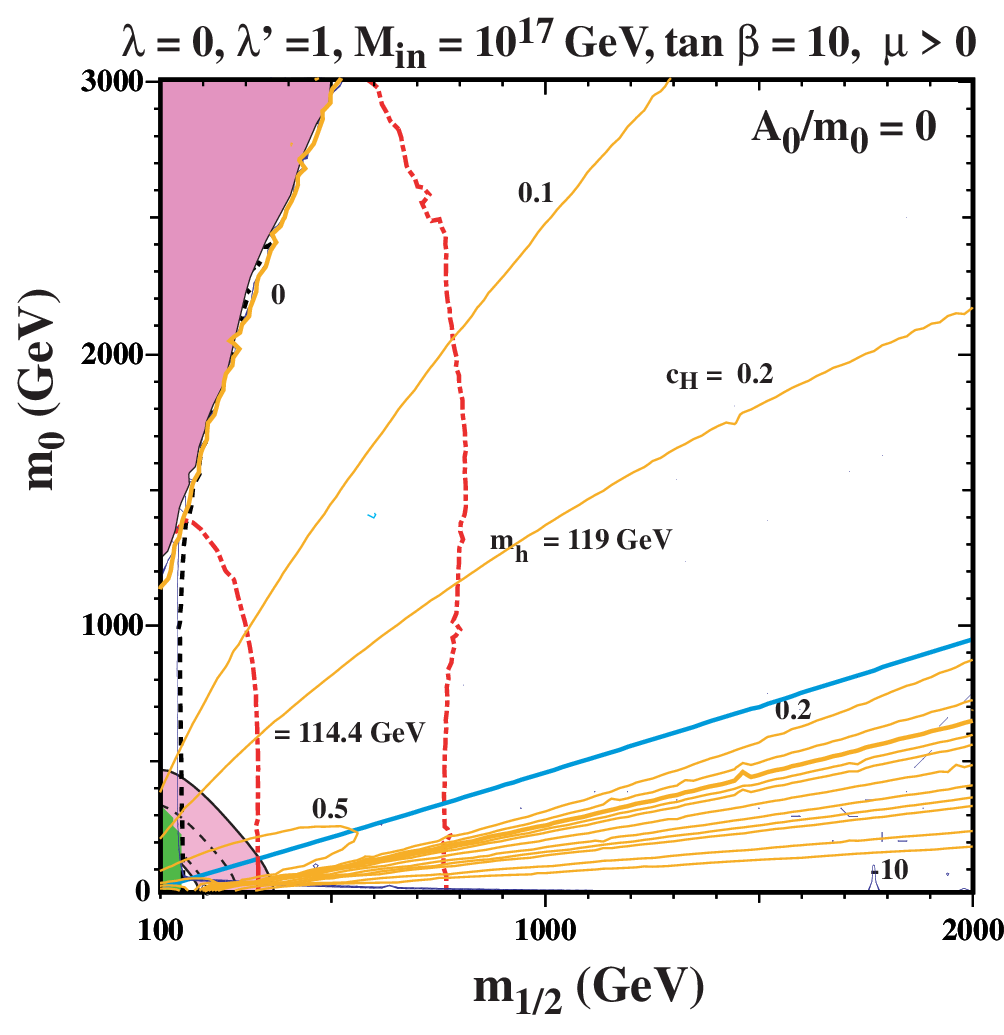,height=8.5cm}
\epsfig{file=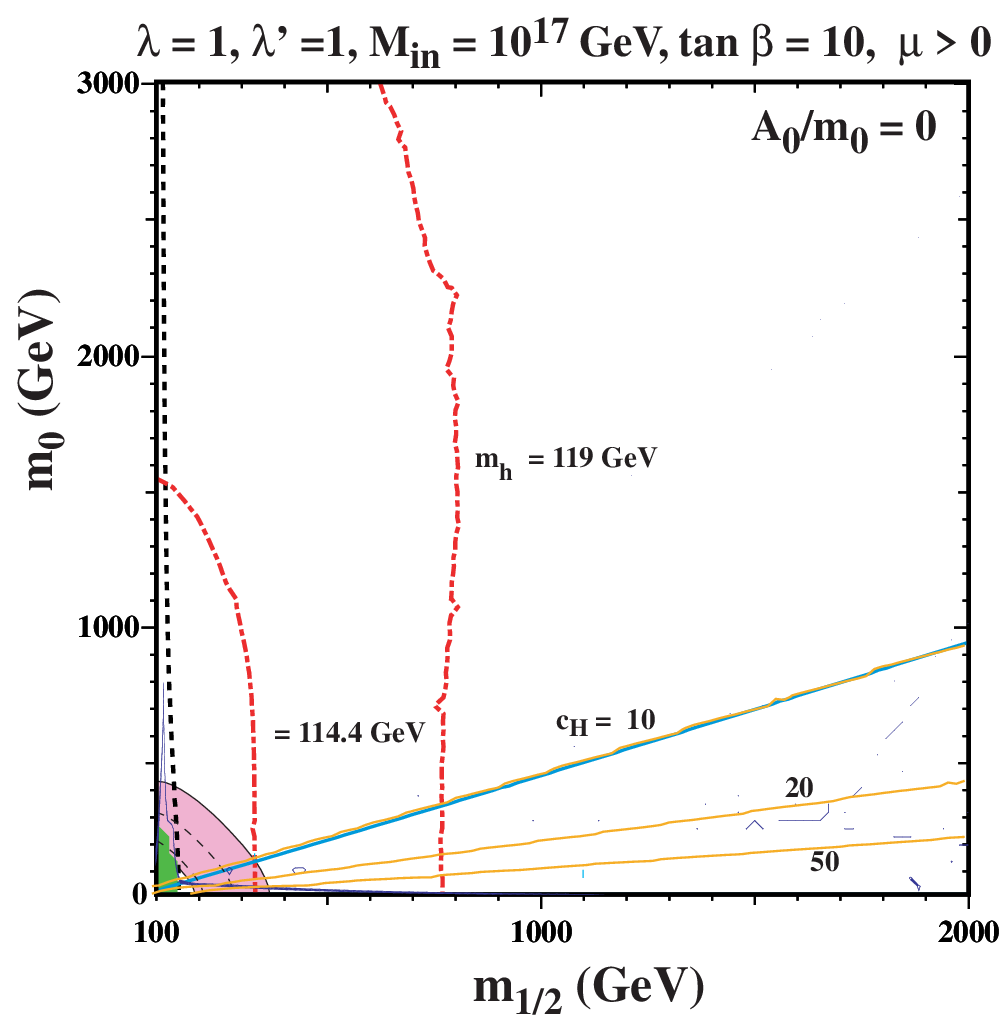,height=8.5cm}\\
\end{center}
\caption{\it
As in Fig. \ref{fig:gmsugra0} with  $\tan \beta = 10$,  $A_0/m_0 = 0$,  $M_{in} = 10^{17}$~GeV,
$\lambda^\prime = 1$, and 
$\lambda = 0$ (left) and  $\lambda = 1$ (right).}
\label{fig:gmsugra17-10}
\end{figure*}

For $\tan \beta = 10$, shown in Fig.~\ref{fig:gmsugra17-10}, 
we see a viable region
for neutralino dark matter only for $\lambda = 0$ along the focus point strip.
(The blue strip here lies under the contour for $c_H = 0$.)  
In fact,  this plane resembles that in Fig.~\ref{fig:msugra0-17}, 
with contours of $\tan \beta$ being replaced by contours of $c_H$ and a different
slope for the Higgs mass contours. 
Values of $c_H$ are acceptable except
at high $m_{1/2}$ and low $m_0$. Notice that unlike the case for $c_H = 0$, the
parameter space does not `close' when $\lambda$ is increased.  In fact, in the
right panel of Fig.~\ref{fig:gmsugra17-10}, we show results for $\lambda = 1$,
a value that would not be possible in mSUGRA (or in the no-scale supergravity \cite{emo2}). 
We see that for $\lambda = 1$ the focus point region is pushed to extremely high values of 
$m_0$ (in excess of 15~TeV). This is due to the additional downward push from the trilinear couplings in the Higgs
mass-squared RGEs that makes $\mu$ larger~\cite{emo}. 
Also values of $c_H$ are significantly higher now: $c_H > 6$ everywhere across the plane
(the gravitino LSP boundary almost coincides with the $c_H$ = 10 contour in this case).

\begin{figure*}
\begin{center}
\epsfig{file=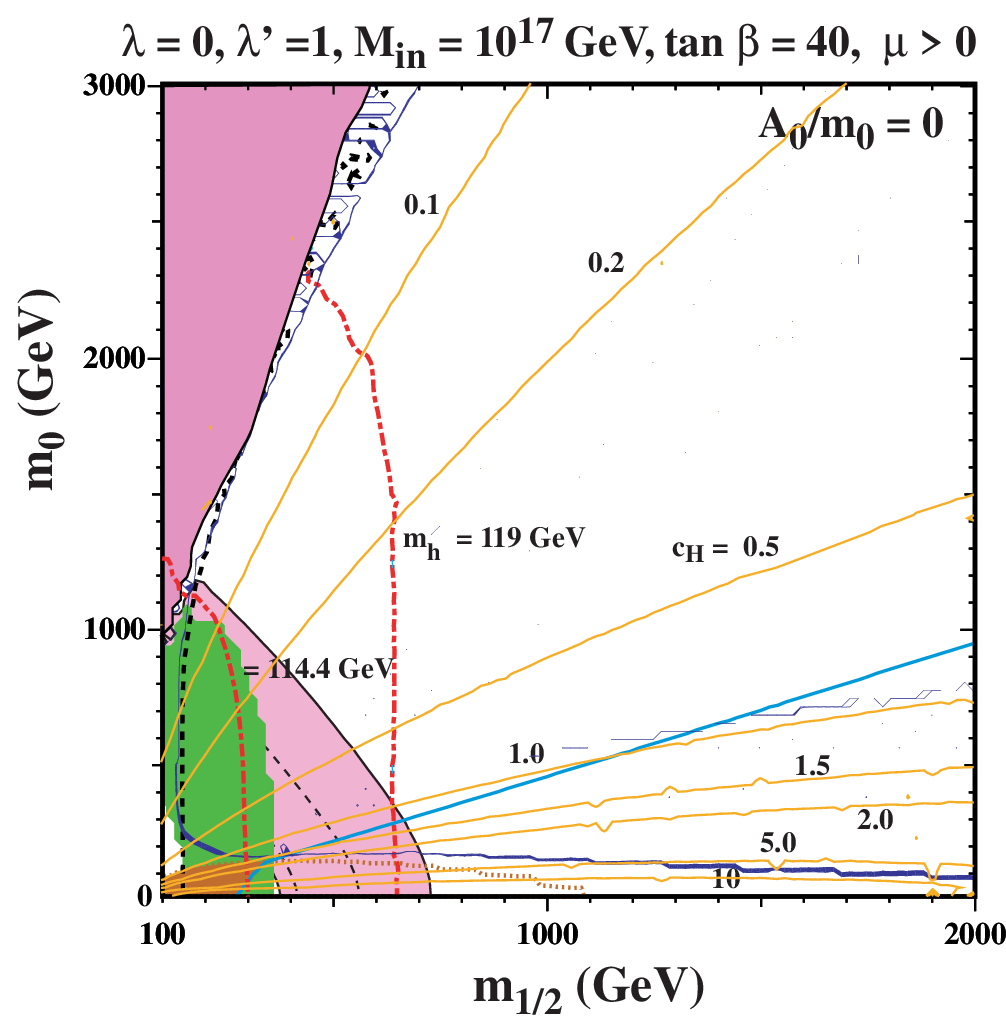,height=8.5cm}
\epsfig{file=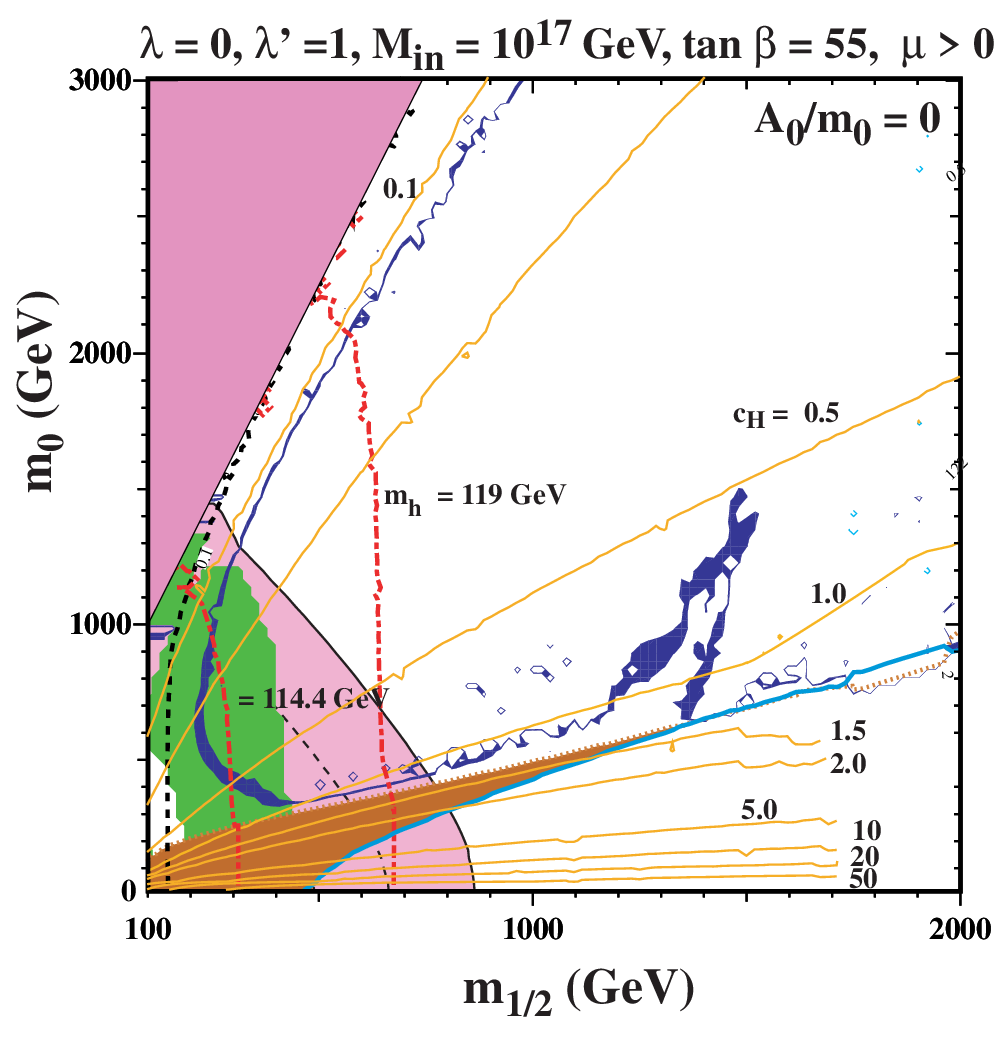,height=8.5cm}\\
\end{center}
\caption{\it
As in Fig. \ref{fig:gmsugra17-10} with  $A_0/m_0 = 0$,  $M_{in} = 10^{17}$~GeV,
$\lambda = 0$, $\lambda^\prime = 1$, for $\tan \beta = 40$ (left) and  $\tan \beta = 55$ (right).  }
\label{fig:gmsugra17-40}
\end{figure*}

In Fig.~\ref{fig:gmsugra17-40}, we show results for $c_H$ for higher values of 
$\tan \beta = 40$ and 55 with $\lambda = 0$ and $A_0 = 0$. For $\tan \beta = 40$, 
the parameter plane is similar to that for $\tan \beta = 10$, with slightly higher values of $c_H$
and a more prominent constraint from $b \to s \gamma$. Again, the focus point strip is the 
only real viable strip for neutralino dark matter. For the larger value of  
$\tan \beta  = 55$, we see the appearance of the rapid annihilation funnel with 
$c_H \sim 1$ and Higgs masses up to 122.5~GeV. The focus point strip is now
clearly seen.  The effect of increasing $\lambda$ can be ascertained from comparing
the left and right panels of Fig.~\ref{fig:gmsugra17-10}. For both $\tan \beta = 40$ and 55,
the focus point region (and the region with no electroweak symmetry breaking) 
will be pushed beyond the scope of the figure for $\lambda \gtrsim 0.5$~\cite{emo}, 
and $c_H$ values will be higher.  
In fact, in both cases, the contours
for $c_H$ will be in roughly the same position as seen in the right panel of Fig.~\ref{fig:gmsugra17-10}.

Also notice a there is a wispy secondary WMAP compatible strip for $\lambda=0$ at  $\m12 \gtrsim 1000$ GeV just above the $c_H = 1$ contour.
Here $m_{\stau_1}\simeq m_A/2$ which enhances $\stau_1$ pair annihilation through $A$ and $H$ Higgs bosons in
the $s-$channel. That enhancement allows one to overcome the suppression 
in stau coannihilation due
to large $\chi-\stau_1$ mass gap and lowers $\ohsq$ to the WMAP range.

\begin{figure*}
\begin{center}
\epsfig{file=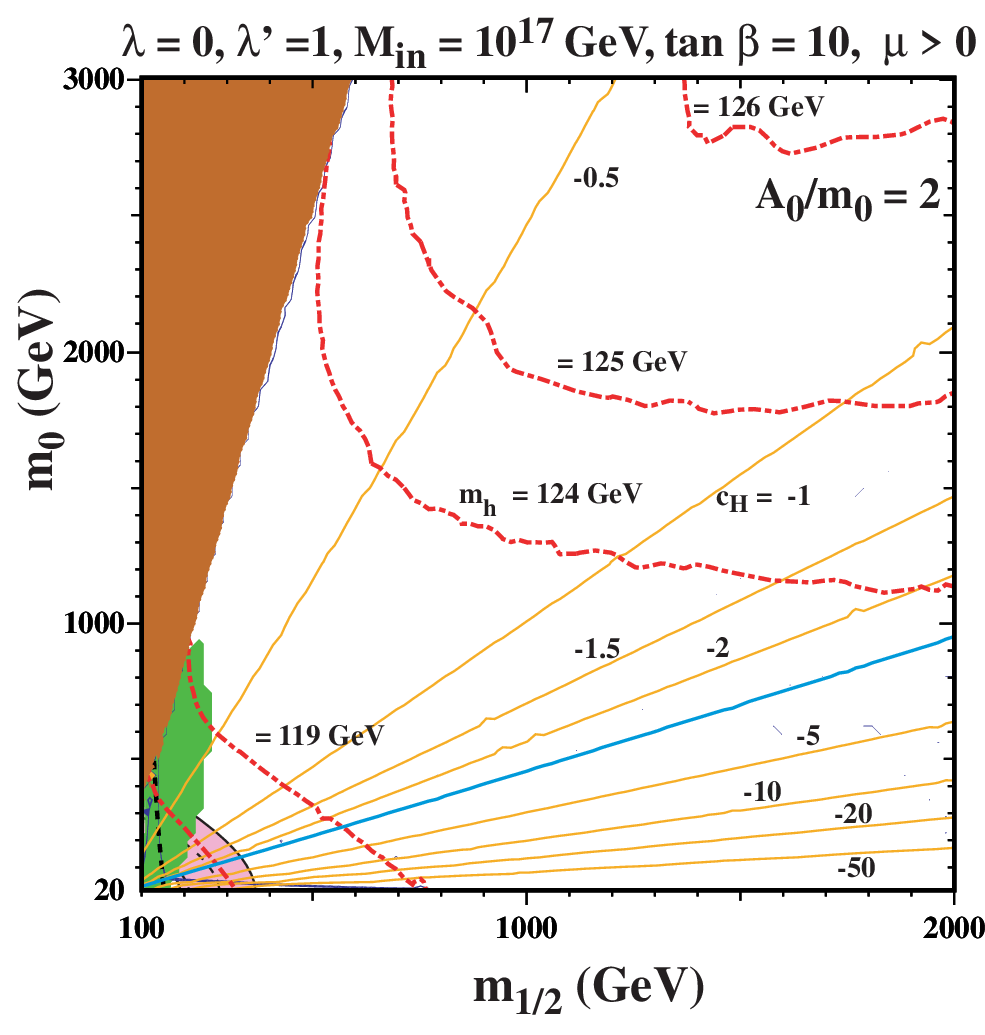,height=8.5cm}
\epsfig{file=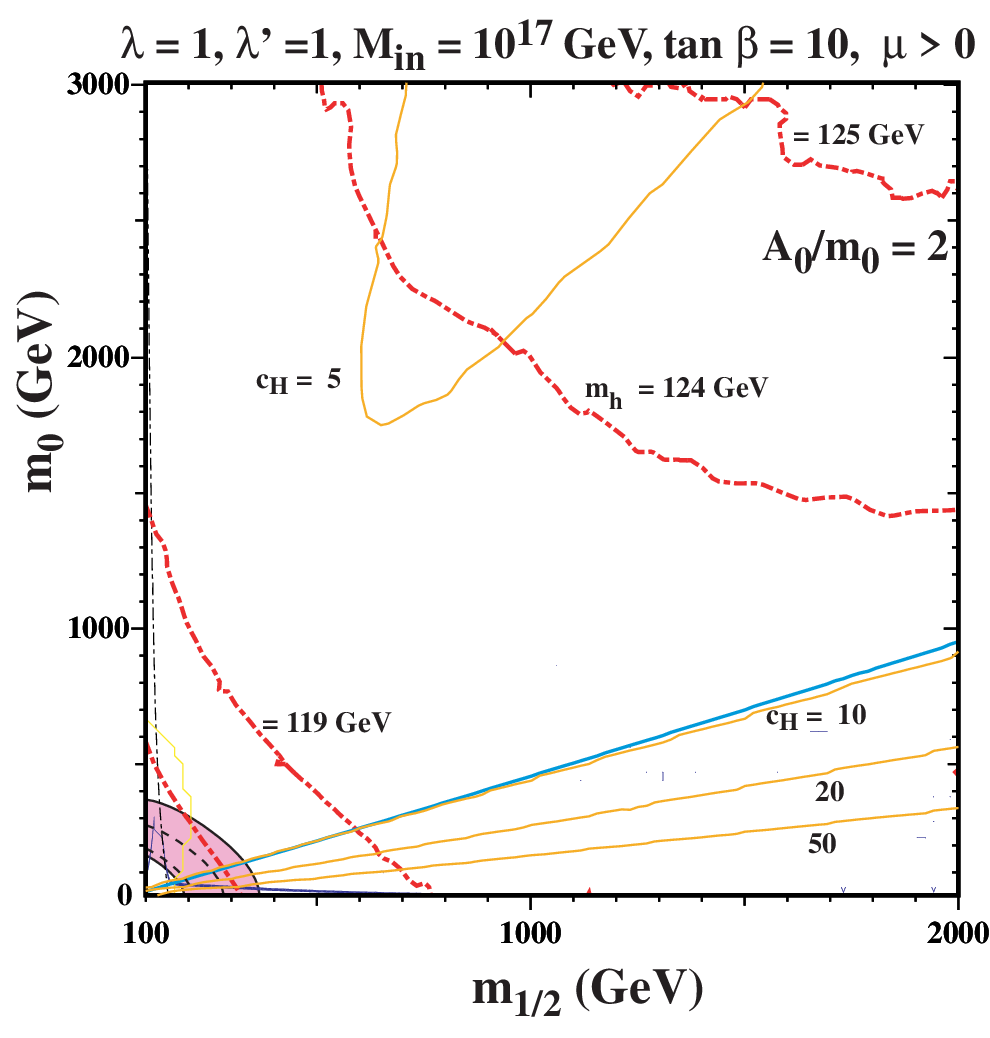,height=8.5cm}\\
\end{center}
\caption{\it
As in Fig. \ref{fig:gmsugra17-10} with  $\tan \beta = 10$,  $A_0/m_0 =2$,  $M_{in} = 10^{17}$~GeV,
$\lambda^\prime = 1$, and 
$\lambda = 0$ (left) and  $\lambda = 1$ (right).}
\label{fig:gmsugra2-17-10}
\end{figure*}

As in the case of GUT scale supergravity, going to higher values of $A_0/m_0$
provides solutions with higher Higgs masses. 
The case for $\tan \beta = 10$ and  $A_0/m_0 = 2.0$ is shown in Fig.~\ref{fig:gmsugra2-17-10}.
Here again, we have a viable stop co-annihilation strip, now with relatively high Higgs masses
$m_h \sim 124$~GeV. For example, for $m_{1/2} \sim 500$~GeV and $m_0 = 2500$~GeV,
$m_{\chi} \sim 240$~GeV and 
one can obtain a neutralino relic density in the WMAP range due to coannihilations with a 
light stop ($m_{\tilde t_1} \sim 270$~GeV).  At this value of $m_0$, all other sfermions
masses are $\gtrsim 1700$~GeV and as a consequence, the contributions to $b \to s \gamma$ and
$B_s \to \mu^+ \mu^-$ are acceptably small. Of course at this value of $m_0$, there is no
way to resolve the $g_\mu-2$ discrepancy.  As seen in the figure $m_h \sim 124$~GeV
and $c_H = -0.36$ at this particular point.

While we are now free to increase $\lambda$ as seen in the right panel of Fig.~\ref{fig:gmsugra2-17-10}, 
where $\lambda = 1$, we see that the $(m_{1/2},m_0)$ plane now looks very 
different. 
A larger value of $\lambda$ causes an increased downward push of the Yukawa terms in the $\aten$ RGE, 
resulting in a smaller value of $A_t$ at $\mgut$. This in turn reduces the downward push in $m^2_{\stop_R}$ 
running below $\mgut$ through the top Yukawa coupling, resulting in a larger value at $M_{weak}$. 
Consequently, $\stop_1$ becomes heavier, for given values of $m_0$ and $\m12$,
and the stop LSP region and the stop co-annihilation strip are moved to higher $m_0$ values beyond the limit of the frame plotted.
In addition, as we have seen before, values of $c_H$ are now significantly higher.

\begin{figure*}
\begin{center}
\epsfig{file=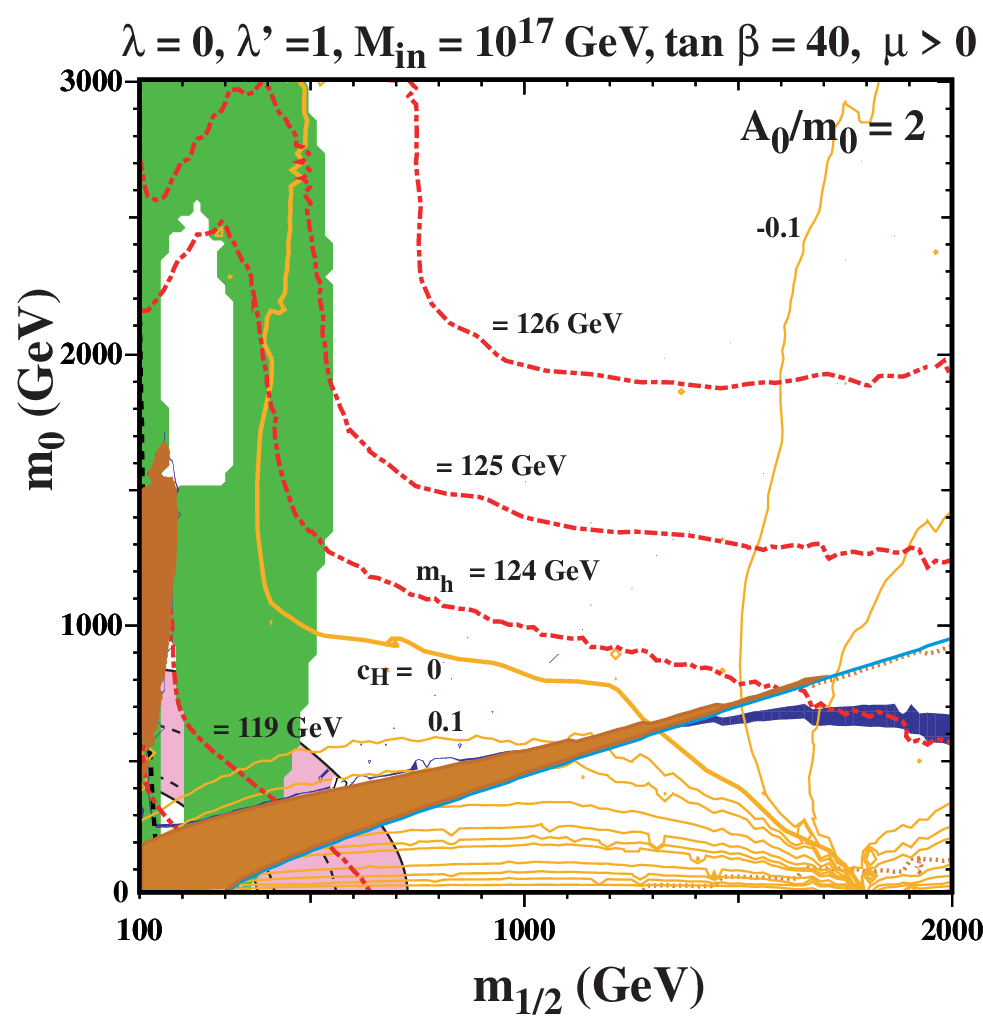,height=8.5cm}
\epsfig{file=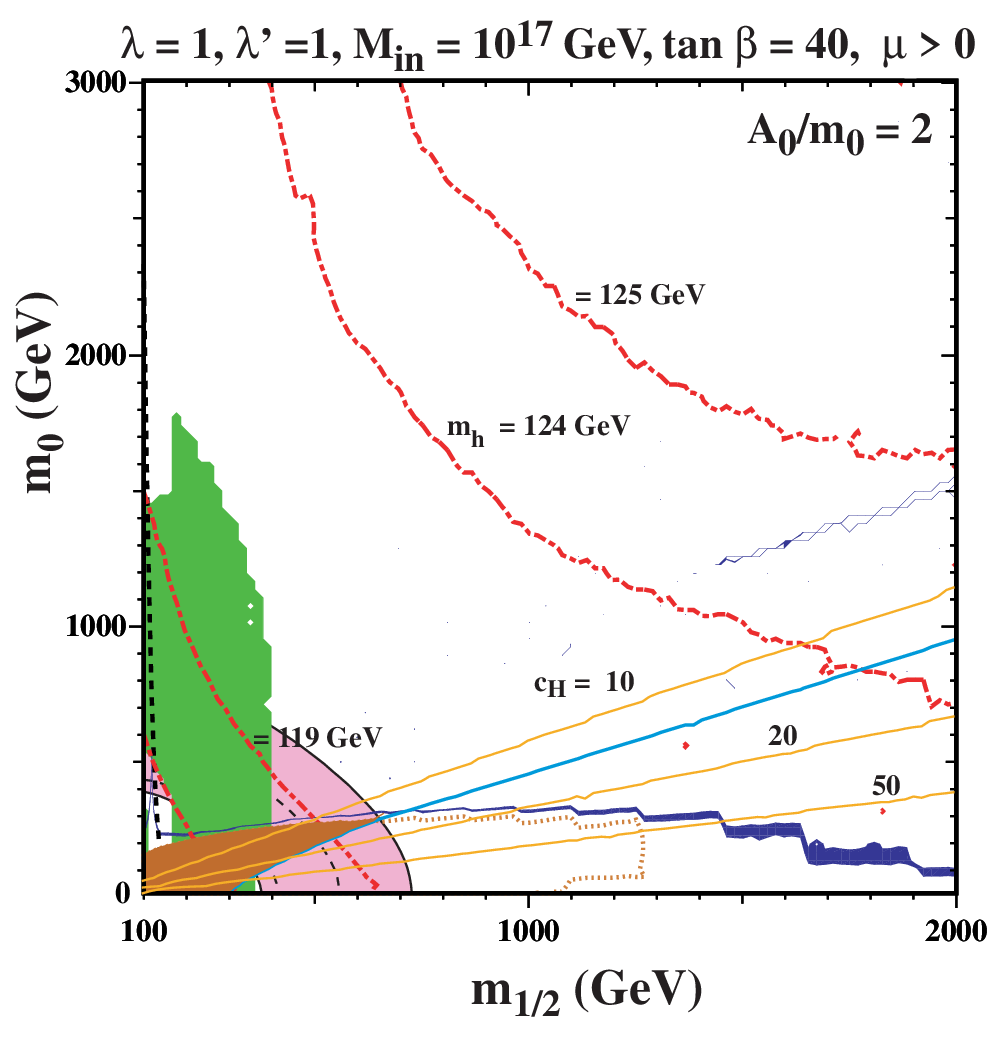,height=8.5cm}\\
\end{center}
\caption{\it
As in Fig. \ref{fig:gmsugra17-10} with  $\tan \beta = 40$,  $A_0/m_0 =2$,  $M_{in} = 10^{17}$~GeV,
$\lambda^\prime = 1$, and 
$\lambda = 0$ (left) and  $\lambda = 1$ (right).  }
\label{fig:gmsugra2-17-40}
\end{figure*}

The plane for $\tan \beta = 40$ is shown in Fig.~\ref{fig:gmsugra2-17-40}.
At first sight, the left panel with $\lambda = 0$,  resembles closely the plane
shown in Fig.~\ref{fig:msugra2-17}.  However, this should not be a surprise as the 
mSUGRA solution for $\tan \beta$ with $M_{in} = 10^{17}$~GeV and $A_0/m_0$ = 2,
is around 40. Indeed, the $c_H = 0$ contour in Fig.~\ref{fig:gmsugra2-17-40} matches
the $\tan \beta = 40$ contour in Fig.~\ref{fig:msugra2-17}. In this case, there are some regions
with neutralino dark matter along the stau co-annihilation strip with $m_h \sim 120$~GeV.
For larger $\lambda = 1$, the co-annihilation strip is somewhat diminished, but 
again, we see that values of $c_H$ are now significantly higher.
For still higher values of $\tan \beta$ with $A_0/m_0 = 2$ and $M_{in} = 10^{17}$~GeV,
we lose the ability to generate sensible spectra (non-tachyonic or neutral LSPs). 
Therefore we do not show the analogous plane for $\tan \beta = 55$.

As in the left panel of Fig.~\ref{fig:gmsugra17-40}, there is also a wispy secondary WMAP compatible strip for $\lambda=1$ at large $\m12$ and $m_0\simeq (1200-1500)$~GeV due to rapid s-channel $\stau_1$ annihilation.

\subsection{$c_H \ne 0, c_\Sigma \ne 0$}
\label{sec:nonzcSig}

The potential problem of large values of $c_H$ seen in the previous subsection
can in principle be alleviated by turning on the second GM parameter, $c_\Sigma$. 
This allows us to more easily satisfy the supergravity boundary conditions
for reasonable values of both $c_H$ and $c_\Sigma$.

One of the reasons that solutions for mSUGRA or no-scale supergravity with $M_{in} > \mgut$  
are only obtained when $\lambda/\lambda^\prime$ is small, is the matching of the $b$-term
in Eq.~(\ref{match}) at $\mgut$.  Since $b$ is dependent on $\tan \beta$, 
and $b_2$ and $\delta b_2$ are fixed by boundary conditions, there is little flexibility in the matching 
condition.  When $\lambda/\lambda^\prime$ is order 1, the contribution from 
$\delta b_2$ is significant and matching at any value of $\tan \beta$ is not guaranteed.
While this problem is alleviated when $c_H \ne 0$, we still have no guarantee that
a particular solution will result in a reasonable value of $c_H$. 
From the definitions in Eq.~(\ref{b2mu2}), we see that in principle, 
a non-zero $c_\Sigma$ can be used to effectively cancel other contributions in $\delta b_2$.
That is we can insure that $\delta b_2$ is small even though $\lambda/\lambda^\prime$ is not.
Of course we have no guarantee that a reasonable value of $c_\Sigma$ can accomplish this
cancellation. 

\begin{figure*}
\begin{center}
\epsfig{file=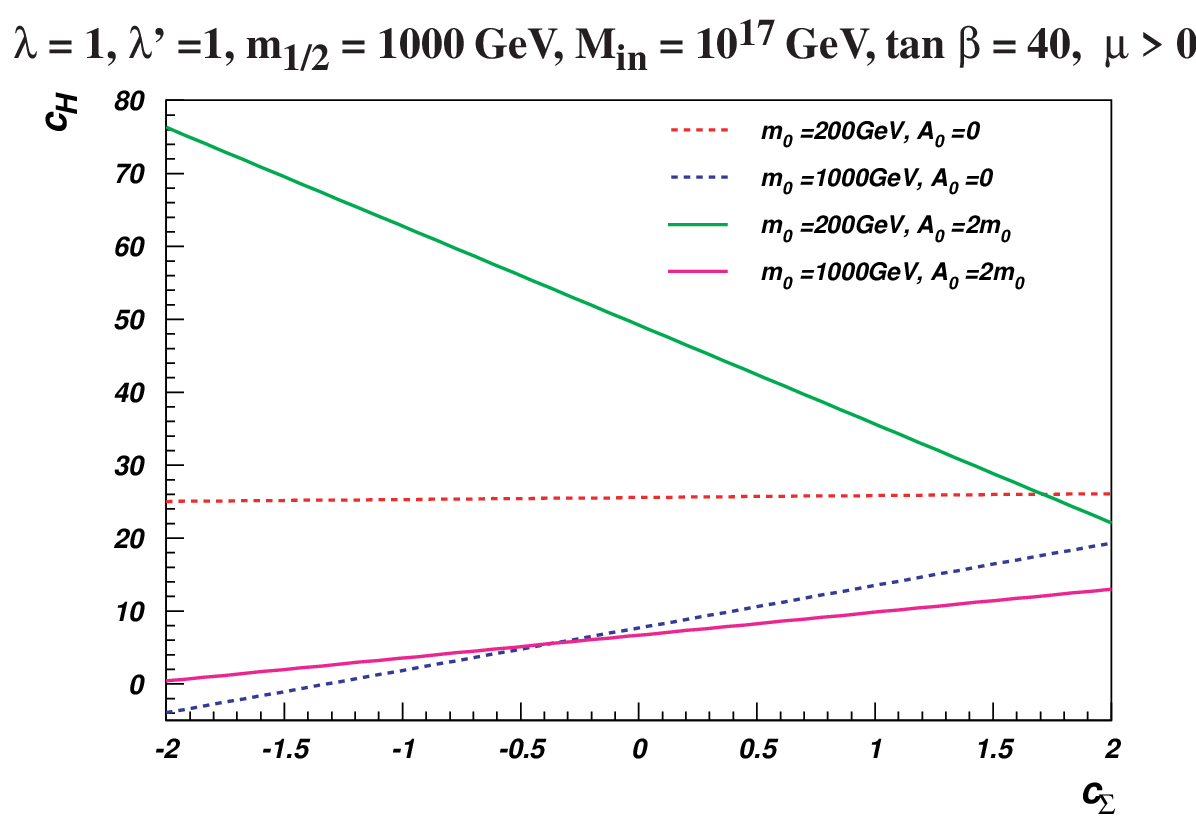,height=5.8cm}
\epsfig{file=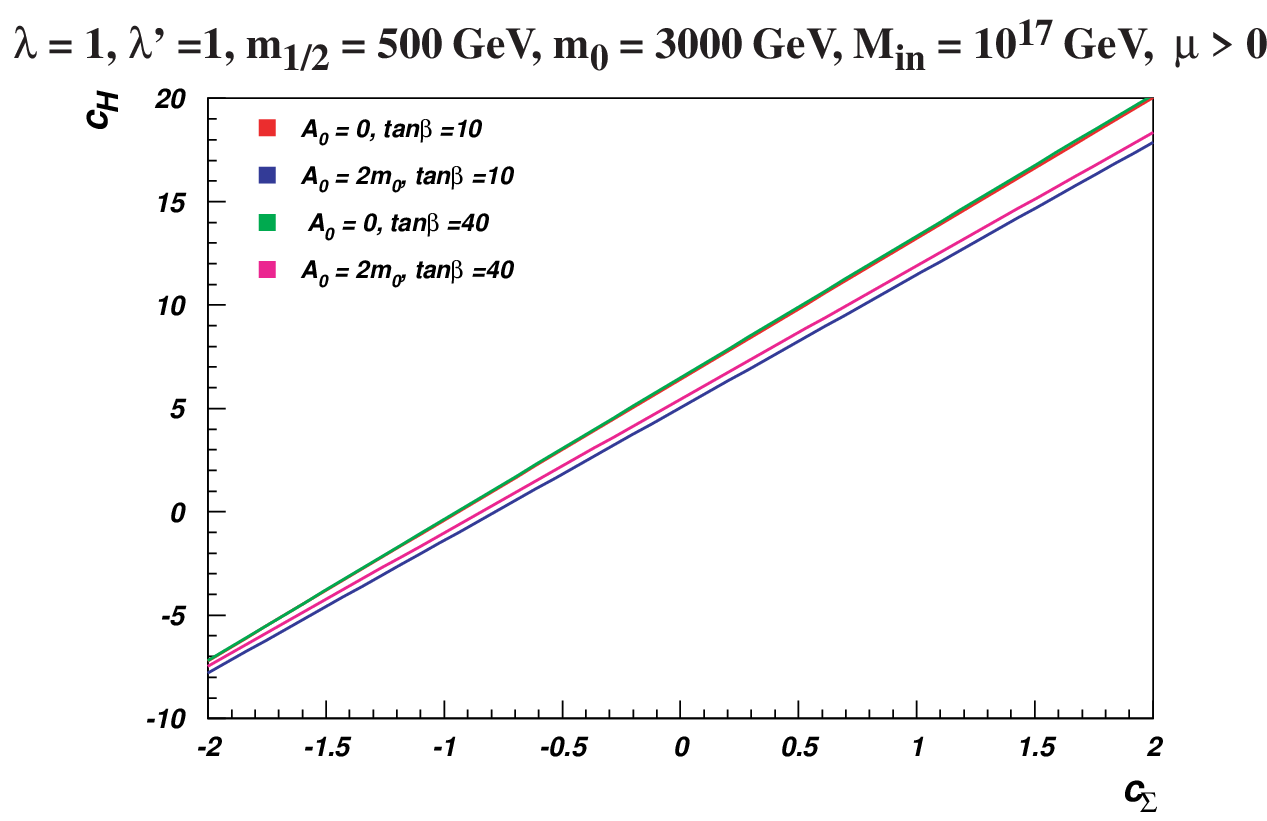,height=5.8cm}
\end{center}
\caption{\it
The resulting value of $c_H$ as a function of an input value of $c_\Sigma$ for several
particular choices of the superGUT parameters. }
\label{fig:csigma}
\end{figure*}

In the left panel of Fig.~\ref{fig:csigma}, we have chosen four points with $m_{1/2} = 1000$~GeV and $\tan \beta = 40$; 
with $m_0 = 200$~GeV and 1000~GeV for $A_0 = 0$ and for $A_0 = 2 m_0$. All four of the points have
relatively high $c_H$ when $c_\Sigma = 0$.  As one can see, the dependence of $c_H$ on
$c_\Sigma$ depends heavily on the particular point. We can get some idea of
what drives this behavior from Eq.~(\ref{ch}).  $c_H$ is proportional to 
${\tilde \mu}_2$ which (at the GUT scale) is $\mu - \delta \mu_2$. The latter is linear in
$c_\Sigma m_0$.   When $m_0$ is small, the change in ${\tilde \mu_2}$ is moderate
and $c_\Sigma$ is determined by the behavior of ${\tilde b}_2/{\tilde \mu}_2$
as compared with $A_0 - m_0$. This could lead to a positive or negative slope.
For the particular cases shown, we see a negative slope when $A_0 = 2 m_0$, and a nearly
flat dependence when $A_0 = 0$.  In contrast, when $m_0$ is large, the effect on ${\tilde \mu}_2$
dominates leading to a positive slope.  In the right panel, we see the relative insensitivity to 
$A_0$ and $\tan \beta$ when $m_0$ is large. Indeed, in these cases, it is possible to 
dial down $c_H$ using reasonably small values of $c_\Sigma$.

\section{Conclusions}
\label{sec:concl}

While often confused in the literature, the CMSSM and mSUGRA are not equivalent
theories nor do they generate the same low energy phenomenology.
The CMSSM is a four-parameter theory (actually five if you include the gravitino mass).
mSUGRA instead is a three-parameter theory. It is also well known that the extra degrees of
freedom in the CMSSM permit the theory to yield a more successful phenomenology,
and in particular, it more easily accommodates the existence of a dark matter candidate with the 
correct relic density \cite{vcmssm,mc4}.

There is however a natural bridge between the two theories.
By minimally extending the K\"ahler potential by including an additional
term of the form given in Eq.~(\ref{gmk})~\cite{gm}, which introduces one new parameter,
one can restore many of the predictions from the CMSSM consistent 
with a UV completion based on supergravity. However in this case (like in mSUGRA),
the gravitino mass remains associated with $m_0$ leaving open the possibility for 
a gravitino dark matter candidate. Here, we have shown that not only is it possible to 
reformulate mSUGRA in this way, but it is possible with reasonably small values
of the new parameter, $c_H$. 
In the case of GUT scale mSUGRA models, regions with an acceptable
neutralino dark matter density are found for relatively large values of $A_0/m_0$.
For $A_0/m_0 = 2$, there is an extended stau co-annihilation strip with $m_h \sim 120$~GeV.
Therefore, most of the promising prediction of the CMSSM
can be recovered in GM supergravity including the possibility of a relatively heavy Higgs boson,
with mass around 125~GeV.

Like the case of no-scale supergravity~\cite{emo2}, in mSUGRA with a superGUT supersymmetry
input scale, the extended running from $M_{in}$ to $\mgut$ makes the phenomenology more difficult, and it is difficult to find solutions for $\tan \beta$.  In essence, $\delta b_2$ in Eq. (\ref{b2mu2})
is relatively large unless the ratio $\lambda/\lambda^\prime$ is small.  For $\lambda = 0$,
solutions for $\tan \beta$ are readily obtained as we saw in Figs.~\ref{fig:msugra0-17} and 
\ref{fig:msugra2-17}. 
At low   $A_0/m_0$, solutions for $\tan \beta$ require a very 
small ratio of the SU(5) Higgs couplings. At higher  $A_0/m_0$, we did find 
solutions for neutralino dark matter along a stop co-annihilation strip but this still required
relatively low $\lambda/\lambda^\prime$. 
Once again, the difficulty in finding solutions for arbitrary $\lambda/\lambda^\prime$
can, in principle, be overcome by introducing a GM extension.  In this case, 
we can add two terms to the K\"ahler potential as in Eq.~(\ref{gmk2}). Adding $c_H$
alone is sufficient for obtaining solutions for arbitrary $\tan \beta$. However, unlike
the preceeding GUT case, here we often find that $c_H$ is large ($\gtrsim 1$)
particularly when $\lambda/\lambda^\prime$ is large.  In some cases, (for example at large $m_0$),
$c_H$ can be tuned down by allowing non-zero $c_\Sigma$.

Ultimately we hope that it will be experiment that sheds light on the viability of these
CMSSM/mSUGRA theories. Here we have tried to explicitly construct a UV completion
to the CMSSM consistent with supergravity in both the case with GUT scale input supersymmetry
breaking and with an input scale above the GUT scale. Despite the extra degree of 
freedom associated with the latter, the additional running from the input scale to $\mgut$
presents some phenomenological challenges.

\section{Acknowledgements}
The work of E.D. was supported in part by the ERC Advanced Investigator Grant no. 226371 ``Mass Hierarchy and Particle Physics at the TeV Scale'' (MassTeV), by the contract PITN-GA-2009-237920. The work of E.D. and Y.M. was supported in part  by the French ANR TAPDMS ANR-09-JCJC-0146.
The work of A.M. and K.A.O. is supported in part by DOE grant DE-FG02-94ER-40823 at the 
University of Minnesota.

\section{Erratum}
\setcounter{equation}{11}

In the original version of the paper, there was an ambiguity between
the value of $\mu$ before and after the shift due to the Giudice-Masiero (GM) term.
Here, we will clarify the equations which were affected.
We define $\mu_0$ as the $\mu$-term in the superpotential defined at the input 
universality scale $M_{in}$.
$\mu(M_{in})$ will refer to the $\mu$-term after the shift induced by the GM contribution to the K\"ahler potential also defined at the input scale.
Then Eq.~(\ref{mushift}) becomes
\beq
\mu(M_{in}) = \mu_0 + c_H m_0 \, .
\label{mushift}
\eeq
Similarly, $\mu B (M_{in})$ is defined as
\beq
\mu B (M_{in}) = \mu_0 B_0 + 2 c_H m_0^2 \, ,
\label{muBshift}
\eeq
which replaces Eq.~(\ref{muBshift}).
As a consequence, we would find
\beq
B(M_{in}) = (A_0 - m_0) \mu_0/\mu(M_{in}) + 2 c_H m_0^2/\mu(M_{in}) \, .
\label{gmb1}
\eeq

\setcounter{equation}{15}
This clarification affects the result only in section 2 of the paper.
For $M_{in} = \mgut$, and when the Giudice-Masiero term (11) is included [15], 
 one can deduce the (GUT) boundary conditions for $\mu$ and $B$
\bea
\mu(\mgut) & = & \mu_0 +  c_H m_0 \, ,
\\
B(\mgut) & = &(A_0-m_0)\mu_0/\mu(\mgut) \nonumber \\
&& +2 c_H m_0^2/\mu(\mgut) \, .
\label{gmb}
\eea
This allows us to solve for $c_H$ where we obtain an equation similar to Eq.~(30)
\beq
c_H = (B(\mgut) - A_0 + m_0)\mu(\mgut)/(3m_0^2 - A_0 m_0) \nonumber \, .
\eeq

These changes affect the contours in Figures 2-4.
In Figure 2, with $A_0 = 0$, all contour labels should be multiplied by 2/3.
In Figure 3, with $A_0 = 2.5 m_0$, all contours should be multiplied by 4.0.
In Figure 4a, with $A_0 = 0$, all contour labels should be multiplied by 2/3.
Finally, in Figure 4b, with $A_0 = 2.0 m_0$, all contour labels should be multiplied by 2.0.

All results and figures in Sections 3 and 4 remain unaffected.

\end{document}